\documentclass[12pt]{amsart}

\usepackage{latexsym, amsmath, amssymb,amsthm, natbib,verbatim, xy}
\xyoption{all}

\usepackage[margin=1.5in]{geometry}

\usepackage{subfig}

\usepackage{tikz}
\usetikzlibrary{shapes}

\usepackage{relsize}

\usepackage{mathtools}

\usepackage{tgpagella}

\usepackage{setspace,color}
\usepackage{graphicx}
\usepackage{enumerate}
\usepackage[multiple]{footmisc}
\usepackage[leqno]{amsmath}
\usepackage{subfig}
\usepackage{hyperref}
\usepackage{xcolor}

\usepackage{epigraph}
\let\originalepigraph\epigraph
\renewcommand\epigraph[2]{\originalepigraph{\textit{#1}}{\textsc{#2}}}

\usepackage{etoolbox}
\patchcmd{\section}{\scshape}{\bfseries}{}{}
\makeatletter
\renewcommand{\@secnumfont}{\bfseries}
\makeatother

\newtheorem{theorem}{Theorem}
\newtheorem{corollary}{Corollary}
\newtheorem{definition}{Definition}
\newtheorem{lemma}{Lemma}
\newtheorem{proposition}{Proposition}

\theoremstyle{remark}

\newtheorem{example}{Example}
\newtheorem*{ex}{Example 1, Continued}

  \def\C{\mathcal{C}}  \def\calc{\mathcal{C}}
    \def\calt{\mathcal{T}}
 \def\cald{\mathcal{D}} 

\def\S{\mathcal{S}} \def\cals{\mathcal{S}}
\def\T{\mathcal{T}} \def\calt{\mathcal{T}}

\newcommand{\df}[1]{\textbf{\textit{#1}}}

\newcommand{\norm}[1]{|| #1 ||}
\newcommand{\abs}[1]{\left| #1 \right|}

\newcommand{\ieh}[1]{{\color{orange} IEH: #1 }}
\newcommand{\fk}[1]{{\color{red} FK: #1 }}

\newcommand{\mby}[1]{{\color{blue} MBY: #1 }}

\begin{document}

\title[Efficient Market Design]{Efficient Market Design with \\ Distributional Objectives$^{\dagger}$}

\author[Hafalir, Kojima, and Yenmez]{Isa E. Hafalir \and Fuhito Kojima  \and M. Bumin Yenmez$^{*}$}

\thanks{
This paper develops and generalizes some of the results in our working paper \cite{interdistrictwp}.
We thank Nanami Aoi, Leo Nonaka, and Ryo Shirakawa for their research assistance and the participants in the 2023 Australasian Economic Theory Workshop for helpful comments. Fuhito Kojima is supported by the JSPS KAKENHI Grant-In-Aid 21H04979. Hafalir is affiliated with the UTS Business School, University of Technology Sydney, Sydney, Australia; Kojima is with the Department of Economics, the University of Tokyo, Tokyo, Japan; Yenmez is with the Department of Economics, Boston College, 140 Commonwealth Ave, Chestnut Hill, MA, 02467. Emails:  \texttt{isa.hafalir@uts.edu.au}, \texttt{fuhitokojima1979@gmail.com},
\texttt{bumin.yenmez@bc.edu}.}


\begin{abstract}
Given an initial matching and a policy objective on the distribution of agent types to institutions,
we study the existence of a mechanism that weakly improves the distributional objective and satisfies constrained efficiency,
individual rationality, and strategy-proofness. We show that such a mechanism need not exist in general. We introduce a
new notion of discrete concavity, which we call pseudo M$^{\natural}$-concavity, and construct a mechanism with the desirable
properties when the distributional objective satisfies this notion. We provide several practically relevant distributional
objectives that are pseudo M$^{\natural}$-concave.
\end{abstract}

\date{\today. First online version: December 31, 2022}

\maketitle



\tikzset{
  every node/.style    = {
    text centered,
    line width = .5,
    anchor     = center,
  },
  every label/.style   = {
    fill   = white,
    anchor = mid,
  },
  every path/.style   = {
    > = stealth
  },
  point/.style   = {
    rounded corners,
    fill           = white,
    minimum height = 20,
    minimum width  = 10,
  },
  rect/.style   = {
    draw,
    minimum height = 12,
    minimum width  = 12,
  },
  triangle/.style   = {
    draw,
    regular polygon,
    regular polygon sides = 3,
    minimum height        = 12,
    minimum width         = 12,
  },
  school/.style   = {
    circle,
  },
  school n/.style   = {
    circle,
    label = { above:#1 },
  },
  school e/.style   = {
    circle,
    label = { right:#1 },
  },
  school s/.style   = {
    circle,
    label = { below:#1 },
  },
  school w/.style   = {
    circle,
    minimum size = 15,
    label = { left:#1 },
  },
  student e/.style   = {
    circle,
    label = { right:#1 },
  },
  student w/.style   = {
    circle,
    label = { right:#1 },
  },
  arrow/.style = {
    ->,
    semithick,
    > = stealth,
  },
  t arrow/.style = {
    ->,
    >          = stealth,
    line width = 3.2,
  },
}


\section{Introduction}
\epigraph{\textit{The political problem of mankind is to combine three things: economic efficiency, social justice, and individual liberty.}}{-John Maynard Keynes (1926)}

In a 1926 lecture, Keynes highlighted the combination of economic efficiency, social justice, and individual liberty as the political problem of mankind.
We study this problem in a matching context, say between institutions and agents of different types. 
In this context, social justice can be interpreted as the policy to improve a \emph{distributional objective} on agent types. Economic efficiency of a matching can be understood as the requirement that there exists no alternative matching improving the welfare of agents without undermining the distributional objective, dubbed as \emph{constrained efficiency}. Individual liberty corresponds to the property that no agent is forced to be matched with an institution against their will, the so-called
\emph{individual rationality}. To Keynes' list, we add \emph{strategy-proofness} as the fourth desideratum because agent preferences need to be elicited to
implement a matching with the desired properties. In this paper, given an initial matching and a distributional objective, we study the existence of a mechanism
that weakly improves the distributional objective and satisfies constrained efficiency, individual rationality, and strategy-proofness.

In many markets, a major goal for policymakers is to improve the distribution of agents to institutions compared to a preexisting matching.
Examples include interdistrict school choice, civil servant matching, teacher assignment, Japanese residency matching, daycare assignment,
and worker and tuition exchange.\footnote{See, \cite{hafalir2022interdistrict} and \cite{kamada2022ekkyo} for
interdistrict school choice; \cite{thakur2021matching} for civil servant matching; \cite{durkes18} and \cite{combe2016design,combe2022market} for teacher
assignment; \cite{kamakoji-basic} for Japanese residency matching; \cite{dur2015two} for worker and tuition exchange; and various other papers in the related literature
section below.} In some applications, agents can be divided
into two sets: newcomers, who are unmatched, and existing agents, who are already matched with institutions. For example, doctors who graduate from medical school
need to be assigned to hospitals. 
Similarly, after teachers get their licenses, they are matched with schools. Many labor markets
share this feature. 
In these markets, the initial matching assigns existing agents with their current institutions, and newcomers are unassigned.
In other applications, agents may have a home institution that defines the initial matching.
For instance, in interdistrict school choice, students are
initially assigned to a school within their district (which may be their neighborhood school) and, in worker exchange, workers are assigned to
the firms they are employed at.
In some of these markets, agent characteristics play an important role and, thus, agents are categorized into different \emph{types}
that can specify, for example, their gender, institution, race, socioeconomic status, disability status, and veteran status.

We define the \emph{distribution} of a matching as a vector that specifies the number of agents of each type at every
institution and the \emph{distributional objective} as a function on distributions that specify how desirable
distributions are in terms of the policy. We seek a mechanism that, among other things, weakly improves the distributional objective
compared to the initial matching. In the rest of the paper, for concreteness, we use the terminology of the
school choice application where students of different types are matched with schools, but our results are applicable
to the aforementioned markets and others as well.

We first observe, for some policy objective on distributions, that there exists no mechanism that weakly improves the distributional objective and
satisfies constrained efficiency, individual rationality, and strategy-proofness (Example \ref{ex:notconvex}). By contrast, we provide a new
mechanism based on the top-trading cycles algorithm of \cite{shasca74} with these properties when the distributional objective is \emph{pseudo M$^{\natural}$-concave},
a notion of concavity for discrete functions that we introduce (Theorem \ref{thm:ttc}).

A distributional objective is pseudo M$^{\natural}$-concave if the minimum value it takes on two distributions is weakly smaller than the minimum value that it takes
when these distributions are made closer to each other. Here, getting closer may either mean adding or subtracting one in a coordinate that we start with,
or the existence of a second coordinate such that we add one in one of the two coordinates and remove one from the other. We show that
a distributional objective is pseudo M$^{\natural}$-concave if and only if its upper contour sets satisfy a notion of discrete convexity called
\emph{M$^{\natural}$-convexity} (Theorem \ref{thm:characterization}). In this sense, pseudo M$^{\natural}$-concavity can be viewed as a
discrete analogue of quasi-concavity. We also show that when there is a set of ideal
distributions and the distributional objective is the negative of the distance to the ideal set
using either the Chebyshev distance or the discrete metric, the distributional objective satisfies
pseudo M$^{\natural}$-concavity if and only if the set of ideal distributions is M$^{\natural}$-convex (Lemmas \ref{lem:cheb} and \ref{lem:policyobjective}).
For instance, if there is a single distribution that is considered ideal, the condition trivially holds.

In the last part of the paper, we provide distributional objectives that are pseudo M$^{\natural}$-concave.
For example, in the worker exchange market, firms may set quota policies where each firm restricts the number of
workers assigned to it between a floor and a ceiling. When students are assigned to schools, each school may have type-specific
floors and ceilings to promote diversity. Furthermore, when there are interdistrict transfers, each district may
want to weakly increase its number of students not to lose any funding. We show that for each of these markets, the distributional objective
corresponding to the policy set is pseudo M$^{\natural}$-concave. We provide additional examples
with pseudo M$^{\natural}$-concave distributional objectives in Section \ref{sec:applications} and Appendix \ref{app:additional}.

\subsection*{Related Literature}\hfill\\
Distributional policies have only recently been studied in the market-design literature. In practice, distributional policies are usually implemented
by reserving seats or positions for target groups in society. \cite{hayeyi13}, \cite{ehayeyi14}, and \cite{echyen12} introduce and study
reserve policies. A prominent example of a matching market with distributional policies is the Japanese Residency Market, where there are constraints on the number
of doctors in regions. \cite{kamakoji-basic,kamakoji-concepts,kamada2018stability,kamada2020accommodating} introduce and study matching markets with
regional constraints such as the Japanese Residency Market. Likewise, distributional policies play an
important role in interdistrict school choice where, historically, students of color
were bused to schools with predominantly white students. \cite{hafalir2022interdistrict} introduce the interdistrict school choice problem and
study distributional policies in this context.\footnote{See a more recent work by \cite{kamada2022ekkyo} as well.} Whereas the main focus of this earlier literature is on finding fair outcomes, we study efficient allocations.



There are only a few papers that study constrained efficient and individually rational mechanisms in the presence of an initial matching and distributional constraints.\footnote{There are other papers that study constrained efficient mechanisms when there are distributional constraints while there is no initial matching. See \cite{abdulson03}, \cite{imamuraefficient}, and \cite{koji2022}. See also \cite{abdulkadirouglu1999house} for efficient mechanisms in a setting with an initial matching and without distributional constraints.} 
\cite{ehayeyi14} consider a setting with type-specific floors and ceilings at schools and construct an algorithm that satisfies constrained efficiency
with other desired properties \citep[Theorem 2]{ehayeyi14}.\footnote{There is no initial matching in the model of \cite{ehayeyi14}, but the
\emph{student-exchange algorithm} that they introduce takes an initial matching as an input and satisfies individual rationality.}
However, their notion of constrained efficiency is different from ours because they also consider a notion of fairness, which makes the comparison
between their work and ours impossible. We consider a distributional objective in their setting and show that a mechanism with the
desired properties exists for this objective (Proposition \ref{prop:convexdiv}). Building on their work,
\citet{suzuki17} study an efficiency notion similar to ours in a market without student types and, therefore, a distribution in their setting specifies only the number of students assigned to each school. Furthermore, their distributional policy is represented by a set of distributions that satisfy it.\footnote{\cite{suzuki2023strategyproof} is an updated journal version of
\citet{suzuki17} where the authors also consider the core
properties of their mechanism, study the deferred acceptance mechanism, and present simulations. \cite{suzuki2023strategyproof}
is subsequent to \cite{interdistrictwp}, which our paper is based on, and contemporaneous to the current work.} They show that a version
of the top trading cycles mechanism satisfies desirable properties if every student is matched initially, the initial matching itself
satisfies the distributional policy, and the set of distributions satisfying the policy goal is M-convex.\footnote{See \citet{kurata2016pareto}
for earlier work in a more specialized setting involving floor constraints at schools.}
While our paper is inspired by their work, there are several crucial differences between our work and \citet{suzuki17}.
First, we allow for student types and, therefore, a distribution in our setting specifies the number of students of each type at every school.
Second, we introduce distributional objectives as functions on distributions and study the existence of
a matching that weakly improves the distributional objective. The motivation for our modeling approach is that, in practice, distributional policies often depend on student attributes
such as socioeconomic status, and they are introduced because the initial matching does not satisfy them. Indeed, policymakers often seek to improve upon the initial situation
toward satisfying the ultimate policy goal, but they do not necessarily insist on satisfying them.\footnote{For example, see the Achievement and Integration Program of the Minnesota Department of Education \citep{hafalir2022interdistrict}.} Another difference is that we allow students to be unmatched.
Nonetheless, we establish a generalization of their main result as Corollary \ref{cor:ttcp}.

Motivated by markets in which institutions swap students (or workers), \cite{dur2015two} study exchange programs and introduce 
algorithms with some desirable properties. A key difference between their work and ours is that, in \cite{dur2015two}, both sides of the market have preferences
and, therefore, their efficiency notion takes into account preferences of both students and institutions. In contrast, we define efficiency only in terms of student preferences.
In addition, the distributional policy of \cite{dur2015two} imposes floors and ceilings on the number of students that an institution may accept,
and this interval includes the number of students that it has before the exchange, whereas we do not focus on any specific policy but consider a general class of
distributional policy objectives. 
However, an application of our results is that in the setting of \cite{dur2015two}, our mechanism satisfies their distributional policy, constrained efficiency, individual rationality,
and strategy-proofness (Proposition \ref{prop:quotas}).

Another related research is \citet{delacretaz2019matching} who study refugee resettlement. They incorporate various constraints into mechanisms similar to the
top-trading cycles algorithm. Constraints studied in their paper are so general that the constraints may fail M$^\natural$-convexity and, therefore, their
mechanisms are not constrained efficient in general. Our study and theirs are complementary in that we identify a class of constraints under
which constrained efficiency can be achieved together with other desiderata while they provide mechanisms that achieve certain efficiency gains over
the initial assignment under general constraints, if not full constrained efficiency.

In a more recent paper, \cite{combe2022market} study how to improve assignments upon the initial matching both in terms of the participants' welfare and distributional objectives similar to our work. However, their model and ours differ in many respects, which makes a direct comparison elusive. One notable difference is that they analyze a highly structured class of distributional objectives, namely those given by a partial order based on the first-order stochastic dominance of the number of students of various types matched at each school, whereas we do not assume such a structure.

The rest of the paper is organized as follows. We present the model in Section \ref{sec:model} and the top trading cycles
mechanism together with its properties in Section \ref{sec:TTC}. We introduce distributional objectives defined in terms of
the distance to an ideal set of distributions in Section \ref{sec:alternative}. We provide distributional objectives from
real-life matching markets that are pseudo M$^{\natural}$-concave in Section \ref{sec:applications}. In Section \ref{sec:conclusion}, we conclude.
We have additional results and all our proofs in the Appendix.

\section{Model}\label{sec:model}
There exist a finite set of schools $\calc$, a finite set of students $\cals$,
and a finite set of student types $\calt$.
Each student $s\in \mathcal{S}$ has a type $\tau_s \in \calt$
and a strict preference relation (a linear order) $P_s$ over $\calc \cup \{\emptyset\}$,
where $\emptyset$ denotes the outside
option of being unmatched.\footnote{Depending on the application, the outside
option can have different interpretations. In the context of school choice, the outside
option can be going to a private school, homeschooling, or moving to a different
city.} The corresponding weak order is denoted by $R_s$.
Each school $c \in \calc$ has a capacity $q_c\in \mathbb{N}$,\footnote{$\mathbb{N}$ represents the set of nonnegative integers.}
which is the maximum number of students who can be enrolled in the school.

A \df{matching} $\mu : \cals \rightarrow \calc \cup \{\emptyset\}$ assigns students to schools
or to the outside option such that no school is assigned more students than its capacity.
In a matching $\mu$, the outcome of student $s\in \cals$ is denoted by
$\mu(s)\in \calc \cup \{\emptyset\}$,
and the outcome of school $c\in \calc$ is denoted by $\mu^{-1}(c)\subseteq \cals$. 
There exists an \df{initial matching} denoted by $\mu_0$. Note that students can be unmatched
at the initial matching. 

A \df{distribution} $\xi \in \mathbb N^{|\C|\times |\T|}$ is a vector such that the entry for school $c\in \calc$ and type $t\in \calt$ is denoted by $\xi_c^t$. The entry
$\xi_c^t$ is interpreted as the number of type-$t$ students at school $c$. Given a
matching $\mu$, the distribution induced by $\mu$ is denoted by $\xi(\mu)$. Formally,
for each school $c\in \calc$ and type $t\in \calt$, $\xi^t_c(\mu)$ is the
number of type-$t$ students assigned to school $c$ in matching $\mu$.
A distribution $\xi \in \mathbb N^{|\C|\times |\T|}$ is \df{feasible} if, for
each school $c\in \calc$, $q_c \geq \sum_{t \in \calt} \xi_c^t$. The set of
feasible distributions is denoted by $\Xi^0$. By definition, $\xi(\mu_0)$ is feasible. 

A \df{distributional objective} $f: \Xi^0 \rightarrow \mathbb{R}$ is a function on distributions such that $f(\xi) \geq f(\xi')$ means that distribution $\xi$ satisfies the objective at least as well as distribution $\xi'$.\footnote{$\mathbb{R}$ represents the set of real numbers. Alternatively, we can define the distributional objectives as
$f: \mathbb N^{|\C|\times |\T|} \rightarrow \mathbb{R} \cup \{-\infty\}$ with $f(\xi)=-\infty$ if and only if $\xi \notin \Xi^0$. Therefore, we only consider feasible distributions.}
Given the initial matching $\mu_0$, a matching $\mu$ \df{weakly improves the distributional objective upon $\mu_0$} if $f(\xi(\mu)) \geq f(\xi(\mu_0))$. When there is no risk of confusion, we simply state that $\mu$ weakly improves the distributional objective.

A matching $\mu$ \df{Pareto dominates} matching $\nu$ if each student weakly prefers her outcome in $\mu$ to her outcome in $\nu$ and,
for at least one student, the preference comparison is strict. Given a distributional objective, a matching $\mu$ that weakly improves
the distributional objective is \df{constrained efficient} if there exists no matching that weakly improves the distributional objective upon $\mu_0$ and
Pareto dominates $\mu$. A matching $\mu$ satisfies \df{individual rationality} if each student weakly prefers her outcome in $\mu$ to her
initial outcome in $\mu_0$, i.e., for each student $s$, $\mu(s)  \mathrel{R_s} \mu_0(s)$.

Throughout the paper, we fix $(\calc, \cals, \calt, (\tau_s)_{s\in \cals}, (q_c)_{c\in \calc}, f, \mu_0)$  and
assume that it is commonly known. Therefore, we refer to a profile of student preferences as a \df{matching problem}. A \df{mechanism} $\phi$ takes a matching problem as
input and produces a matching. The outcome for student $s\in \mathcal{S}$ at the reported preference
profile $P$ under mechanism $\phi$ is denoted as $\phi_s(P)$. A mechanism $\phi$ satisfies \df{strategy-proofness}
if no student can misreport her preferences and get a strictly more preferred outcome. More
formally, for each student $s\in \cals$ and preference profile $P$, there exists no
preference relation $P'_s$ such that $\phi_s(P'_s,P_{-s}) \mathrel{P_s} \phi_s(P)$,
where $P_{-s}$ denotes the preference profile of students excluding $s$.

For any property on matchings (such as constrained efficiency and individual rationality),
a mechanism satisfies the property if, for each preference profile, the matching produced by
the mechanism satisfies the property at the preference profile.

\section{Improving the Distributional Objective}\label{sec:TTC}
We study the existence of a mechanism that weakly improves the distributional objective and satisfies
constrained efficiency, strategy-proofness, and individual rationality.

\subsection{An Example}\hfill\\
We first show that, for some distributional policies, there exists no mechanism
with the desirable properties. The following example establishes this result.

\begin{example}\label{ex:notconvex}
Suppose that the set of schools is $\{c_1, \ldots ,c_6\}$,
the set of students is $\{s_1, \ldots, s_6\}$, and the set of student
types is $\{t_1,t_2,t_3\}$. Each school has a capacity of one.
Students $s_1$ and $s_4$ have type $t_1$, students $s_2$ and $s_5$ have type $t_2$, and students $s_3$ and $s_6$ have type $t_3$.
The distributional policy objective is given as, for every feasible distribution $\xi\in \Xi^0$,
\[  f(\xi) = \begin{cases*}
                    1 & if $\sum_{t,c} \xi_c^t = 6$ and $\sum_{c \in \{c_1, c_2, c_3\}} \xi_c^t = 1$ for each $t\in \calt$; and  \\
                    0 & otherwise.
                 \end{cases*} \]
In the initial matching, student $s_i$ is matched with school $c_i$, where $i=1,\ldots,6$. Therefore, the initial matching is such that all six students are assigned to a school and, for each type $t\in \calt$, there is one type-$t$ student assigned to a school in $\{c_1,c_2,c_3\}$. Therefore, $f(\xi(\mu_0)) = 1$.

Student preferences are as follows:
\[
\begin{tabular}
[c]{llllll}%
$\underline{s_{1}}$ & $\underline{s_{2}}$ & $\underline{s_{3}}$ &
$\underline{s_{4}}$ & $\underline{s_{5}}$ & $\underline{s_{6}}$\\
$c_6$ & $c_6$ & $c_5$ & $c_{3}$ & $c_3$ & $c_1$\\
$c_1$ & $c_2$ & $c_4$ & $c_4$ & $c_{5}$ & $c_2$\\
\vdots & \vdots & $c_3$ &  \vdots & \vdots  & $c_6$\\
          &            & \vdots &             &            & \vdots
\end{tabular}
\]
\smallskip

\noindent
where the dots in the table mean that the corresponding parts of the preferences over the schools are arbitrary, and being unassigned is the least preferred option for each student.

In this example, there are two matchings that weakly improve the distributional objective and
satisfy constrained efficiency and individual rationality:
\begin{align*}
\mu & =\{(  s_{1},c_6)  ,(s_{2},c_2)
,(s_{3},c_{4}),(s_{4},c_{3}),(s_{5}
,c_{5})  ,(s_{6},c_1)\}, \text{ and} \\
\tilde \mu  & =\{(s_{1},c_{1}),(s_{2}
,c_6),(s_{3},c_5),(s_{4},c_4)
,(s_{5},c_3),(s_{6},c_2)\}.
\end{align*}
Therefore, if a mechanism satisfies the desired properties, then its outcome at the above student preference profile must be either matching $\mu$ or $\tilde \mu$.

Consider the case where the mechanism produces matching $\mu$ at the above student preference profile. Suppose student $s_3$ misreports her preference by ranking $c_5$ first and $c_3$ second while the ranking of other schools is arbitrary. Under the new report, the mechanism produces matching $\tilde \mu$ because it is the only matching that weakly improves the distributional objective and satisfies constrained efficiency and individual rationality. Since student $s_3$ strictly prefers her school at $\tilde \mu$ to her school at $\mu$, she has a profitable deviation.

Similarly, consider the case where the mechanism produces matching $\tilde \mu$ at the above student preference profile. Suppose student $s_6$ misreports her preference by ranking $c_1$ first and $c_6$ second while the ranking of the other schools is arbitrary. In this case, the mechanism produces matching $\mu$ because it is the only matching that weakly improves the distributional objective
and satisfies constrained efficiency and individual rationality. Since student $s_6$ strictly prefers her school at $\mu$ to her school at $\tilde \mu$, she has a profitable deviation.

In both cases, there exists a student who benefits from misreporting her preferences, so there exists no mechanism with the desirable properties.
\qed
\end{example}

Hence, for a given distributional objective, it may not be possible to achieve the desired properties.

\subsection{Pseudo M$^{\natural}$-concavity}\hfill\\
To achieve the existence of a mechanism with the desirable properties, we introduce a new notion of discrete
concavity and consider distributional objectives that satisfy it.

Let $\chi_c^t \in \mathbb N^{|\C|\times |\T|}$ denote
the distribution where there is one student of type $t\in \mathcal{T}$ at school $c\in \calc$ and there are no other students.

\begin{definition}\label{def:pmc}
A distributional objective $f$ is \textbf{pseudo M$^{\natural}$-concave} if, for every feasible distributions
$\xi,\tilde{\xi} \in \Xi^0$ and $(c,t) \in \calc \times \calt$ with $\xi_c^t>\tilde{\xi}_c^t$, either
\begin{enumerate}
\item $\min \{f(\xi-\chi_c^t), f(\tilde{\xi}+\chi_c^t)\} \geq \min \{f(\xi),f(\tilde{\xi})\}$ or
\item there exists $(c',t') \in \calc \times \calt$ with $\xi_{c'}^{t'}<\tilde{\xi}_{c'}^{t'}$ such that
\begin{equation*}
  \min \{f(\xi-\chi_c^t+\chi_{c'}^{t'}), f(\tilde{\xi}+\chi_c^t-\chi_{c'}^{t'})\} \geq \min \{f(\xi),f(\tilde{\xi})\}.
\end{equation*}
\end{enumerate}
\end{definition}
Intuitively, pseudo M$^{\natural}$-concavity requires that when we move from two distributions towards each other, the minimum value
of the distributional objective does not decrease. The definition not only requires the stated inequalities but also that the arguments of
the function are feasible. As will be shown in Theorem \ref{thm:characterization}, pseudo M$^{\natural}$-concavity is analogous to quasi-concavity.

Let $\chi_{\emptyset}^{\emptyset} \in \mathbb N^{|\C|\times |\T|}$ be the
zero vector for notational convenience. Using this notation, pseudo M$^{\natural}$-concavity can be written as follows:
A distributional objective $f$ is pseudo M$^{\natural}$-concave if, for every feasible distributions
$\xi,\tilde{\xi} \in \Xi^0$ and $(c,t) \in \calc \times \calt$ with $\xi_c^t>\tilde{\xi}_c^t$, there exists
$(c',t') \in (\calc \times \calt)\cup \{(\emptyset,\emptyset)\}$ (with $\xi_{c'}^{t'}<\tilde{\xi}_{c'}^{t'}$ whenever $(c',t')\in \calc \times \calt$) such that
\begin{equation*}
  \min \{f(\xi-\chi_c^t+\chi_{c'}^{t'}), f(\tilde{\xi}+\chi_c^t-\chi_{c'}^{t'})\} \geq \min \{f(\xi),f(\tilde{\xi})\}.
\end{equation*}
We use this notation in the rest of the paper.

Pseudo M$^{\natural}$-concavity is weaker than the following notion.
A distributional objective $f$ is \textbf{pseudo M-concave} if, for every feasible distributions
$\xi, \tilde{\xi} \in \Xi^0$ and $(c,t) \in \calc \times \calt$ with $\xi_c^t>\tilde{\xi}_c^t$,
there exists $(c',t') \in \calc \times \calt$  with $\xi_{c'}^{t'}<\tilde{\xi}_{c'}^{t'}$ such that
\[\min \{f(\xi-\chi_c^t+\chi_{c'}^{t'}), f(\tilde{\xi}+\chi_c^t-\chi_{c'}^{t'})\} \geq \min \{f(\xi),f(\tilde{\xi})\}.\]

The failure of a mechanism with the desirable properties in Example \ref{ex:notconvex} can be
attributed to the distributional objective failing pseudo M$^{\natural}$-concavity.

\begin{ex}
Recall that matchings $\mu$ and $\tilde \mu$
are such that $f(\xi(\mu))=f(\xi(\tilde \mu))=1$. Denote $\xi(\mu)$ by $\xi$ and $\xi(\tilde \mu)$ by $\tilde{\xi}$. Observe that $\xi_{c_3}^{t_1}=1>0=\tilde{\xi}_{c_3}^{t_1}$ because
(i) school $c_3$ is matched with student $s_4$ at $\mu$, whose type is $t_1$ and (ii) school $c_3$ is matched only
with student $s_5$ at $\tilde \mu$, whose type is $t_2$.

We show that $f$ is not pseudo M$^{\natural}$-concave. Towards a contradiction, suppose that $f$ satisfies the condition.
Since $\xi_{c_3}^{t_1}>\tilde{\xi}_{c_3}^{t_1}$, pseudo M$^{\natural}$-concavity implies
the existence of $(c',t') \in (\calc \times \calt)\cup \{(\emptyset,\emptyset)\}$
(with $\xi_{c'}^{t'}<\tilde{\xi}_{c'}^{t'}$ whenever $(c',t') \in \calc \times \calt$) such that
\begin{equation*}
  \min \{f(\xi-\chi_{c_3}^{t_1}+\chi_{c'}^{t'}), f(\tilde{\xi}+\chi_{c_3}^{t_1}-\chi_{c'}^{t'})\} \geq \min \{f(\xi),f(\tilde{\xi})\}.
\end{equation*}
Note that, $f(\xi) = f(\tilde{\xi}) = 1$ and, furthermore,  $f(\xi-\chi_{c_3}^{t_1}) = 0$ because
not all students are matched at $\xi-\chi_{c_3}^{t_1}$.
Therefore, we need to have $(c',t') \in \calc \times \calt$ (equivalently, $(c',t') \neq (\emptyset,\emptyset)$) for the displayed inequality to hold.
Furthermore, we need a school $c'\in \calc$ and a type $t'\in \calt$ such that
$\xi_{c'}^{t'}<\tilde{\xi}_{c'}^{t'}$ and $f(\xi-\chi_{c_3}^{t_1}+\chi_{c'}^{t'}) = 1$. This requires $\xi-\chi_{c_3}^{t_1}+\chi_{c'}^{t'}$ to be feasible where
all students are matched, and hence the only candidate for $(c',t')$ satisfying the above condition is
when $c'=c_3$. Next, $\xi_{c_3}^{t'}<\tilde{\xi}_{c_3}^{t'}$ requires $\tilde{\xi}_{c_3}^t$ be  strictly positive, and the only non-zero $\tilde{\xi}_{c_3}^t$ is for $t=t_2$
(because $s_5$ is the unique student matched with $c_3$ at $\tilde \mu$). Finally, $f(\xi-\chi_{c_3}^{t_1}+\chi_{c_3}^{t_2})=0$
gives a contradiction to the displayed inequality.

We conclude that $f$ is not pseudo M$^{\natural}$-concave.
\qed
\end{ex}

\subsection{Top Trading Cycles Mechanism}\label{ttc_sec}\hfill\\
We now introduce a mechanism that achieves the desirable properties 
whenever the distributional objective is pseudo M$^{\natural}$-concave.

We first create a hypothetical two-sided matching market. On one side of the market, there are school-type pairs $(c,t) \in \C \times \calt$, and also an outside option $(\emptyset,\emptyset)$ that represents being unmatched. On the other side, there are students from the original market, $\S$.

Given any student $s \in \S$ with type $t\in \calt$ and her preference relation $P_s$ in the original market, let $\tilde P_s$ be a preference relation over
school-type pairs and the outside option in the hypothetical market such that, for every $c,c' \in \C$ and $t'\in \calt \setminus \{t\}$, we have,
\begin{enumerate}[(i)]
\item $c \mathrel{P_s} c' \iff (c,t) \mathrel{\tilde P_s} (c', t)$,
\item $c \mathrel{P_s} \emptyset \iff (c,t) \mathrel{\tilde P_s} (\emptyset,\emptyset)$,
\item if $\mu_0(s) \in \mathcal{C}$, then $(\mu_0(s),t) \mathrel{\tilde P_s} (c,t')$, and
\item if $\mu_0(s) = \emptyset$, then $(\emptyset,\emptyset) \mathrel{\tilde P_s} (c,t')$.
\end{enumerate}
That is, $\tilde P_s$ is a preference relation over school-type pairs and the outside option that ranks the school-type
pairs in which the type is $t$ in the same order as in $P_s$, while finding all school-type
pairs specifying a different type as less preferred than the pair corresponding to her initial outcome.
Let $\mathrel{\tilde R_s}$ denote the weak preference corresponding to $\mathrel{\tilde P_s}$.

Next, we define a priority ordering of students that school-type pairs and $(\emptyset,\emptyset)$ use to rank students. For a pair
$(c,t)\in \calc \times \calt$, students initially matched with $(c,t)$ (i.e., $\mu_0(s)=c$ with $\tau_s = t$) have the highest priority, and then all other students have
the second highest priority. For $(\emptyset,\emptyset)$, students who are initially unmatched have the highest priority, and then all
other students have the second highest priority. This gives us two priority classes for students at each school-type pair and $(\emptyset,\emptyset)$.
We break ties within each class according to a master priority list over students that every school-type pair and $(\emptyset,\emptyset)$ uses.

Now we define the set of students that each school-type pair or $(\emptyset,\emptyset)$ can point to in our algorithm below.
At a matching $\mu$, a student $s\in \cals$ with type $t\in \calt$ and $\mu_0(s)=c\in \calc$ is \df{permissible}
to $(c',t') \in \left(\calc \times \calt\right) \cup \{(\emptyset,\emptyset)\}$ 
if \[f(\xi(\mu)+\chi_{c'}^{t'}-\chi_c^t) \geq f(\xi(\mu_0)).\]
Note that such a student is permissible to pair $(c,t)$ at a matching $\mu$ whenever $f(\xi(\mu)) \geq f(\xi(\mu_0))$.
Similarly, at a matching $\mu$, a student $s\in \cals$ with type $t\in \calt$ and $\mu_0(s)=\emptyset$ is \df{permissible}
to $(c',t')\in \left(\calc \times \calt\right) \cup \{(\emptyset,\emptyset)\}$ 
if \[f(\xi(\mu)+\chi_{c'}^{t'}) \geq f(\xi(\mu_0)).\]

We generalize the top trading cycles algorithm of \citet{shasca74} to our setting as follows.
\medskip

\paragraph{\textbf{Top Trading Cycles (TTC) Algorithm}}
Consider a hypothetical market corresponding to a matching problem.
\begin{description}
  \item[Step 1] Let $\mu^1 \equiv \mu_0$. Each school-type pair and $(\emptyset,\emptyset)$ points to the permissible student at matching $\mu^1$ with the highest priority.
  If there exists no such student, remove the school-type pair or $(\emptyset,\emptyset)$ from the market. Each student $s$ points to the highest ranked
  remaining school-type pair or $(\emptyset,\emptyset)$ with respect to $\tilde{P}_s$. Identify and execute cycles by assigning each student the school-type pair or $(\emptyset,\emptyset)$ she is pointing to. Remove all students in the identified cycles from the market.
  \item[Step $\mathbf{n}$ ($\mathbf{n>1}$)] Let $\mu^n$ denote the matching consisting of assignments in the previous steps and initial assignments for all students who have not been processed in the previous steps. Each remaining school-type pair and $(\emptyset,\emptyset)$ points to the student who is permissible at matching $\mu^n$ with the highest priority. If there exists no such student, remove the school-type pair or $(\emptyset,\emptyset)$ from the market. Each student $s$ points to the highest-ranked remaining school-type pair or $(\emptyset,\emptyset)$ with respect to $\tilde{P}_s$. Identify and execute cycles by assigning each student the school-type pair or $(\emptyset,\emptyset)$ she is pointing to. Remove all students in the identified cycles from the market.
\end{description}

The algorithm terminates when no student remains to be processed. The outcome is defined as the matching corresponding to
the outcome of the hypothetical market at the final step in the following sense. A student $s\in \cals$ is assigned to school $c\in \calc$
at the outcome of the TTC algorithm if $s$ is assigned to $(c,t)$, for some $t\in \calt$, in the hypothetical market and $s$ is unassigned
at the outcome of the TTC algorithm if $s$ is assigned to $(\emptyset,\emptyset)$ in the hypothetical market. We provide an illustrative
example of the TTC algorithm in Appendix \ref{subsec:ttcexample}. Note that the definition of permissibility and, hence, the definition of
the TTC algorithm depends on the distributional objective $f$. Nevertheless, we do not explicitly state the distributional objective under
consideration when it is clear from the context.

The \textbf{top trading cycles (TTC) mechanism} takes a profile of student preferences as input and produces the outcome of the TTC algorithm
at the reported student preference profile.

Our main result is as follows.

\begin{theorem}\label{thm:ttc}
Suppose that the distributional objective $f$ is pseudo M$^{\natural}$-concave. Then the TTC mechanism weakly improves the distributional objective and satisfies constrained efficiency, individual rationality, and strategy-proofness.
\end{theorem}

\section{Sets of Distributions as Policy Goals}\label{sec:alternative}
In this section, we study markets in which policymakers identify their policy goal with
a set of ideal distributions and their distributional objective is defined using this set.\footnote{We provide examples of such policies
in Section \ref{sec:applications} and in Appendix \ref{app:additional}.}
Therefore, we represent a policy goal as a non-empty set of distributions $\Xi\subseteq \Xi^0$ and
say that a matching $\mu$ \df{satisfies the policy goal} $\Xi$ if the distribution induced by $\mu$ is in $\Xi$, that is, $\xi(\mu) \in \Xi$.

\subsection{Discrete Convexity}\label{sec:convexity}\hfill\\
We use the following notion of discrete convexity on policy goals, which is studied in
mathematics and operations research literatures \citep{Murota:SIAM:2003}.

\begin{definition}\label{def:mnat}
A policy goal $\Xi\subseteq \Xi^0$ is M$^{\natural}$\textbf{-convex} if,
for every $\xi,\tilde{\xi} \in \Xi$ and $(c,t) \in \calc \times \calt$ with $\xi_c^t>\tilde{\xi}_c^t$,
there exists $(c',t') \in \left( \calc \times \calt \right) \cup \{(\emptyset,\emptyset)\}$
(with $\xi_{c'}^{t'}<\tilde{\xi}_{c'}^{t'}$ whenever $(c',t')\in \calc \times \calt$) such that
\[\xi-\chi_c^t+\chi_{c'}^{t'}\in \Xi \; \mbox{ and } \; \tilde{\xi}+\chi_c^t-\chi_{c'}^{t'} \in \Xi.\]
\end{definition}
For example, $\Xi^0$ is M$^{\natural}$-convex. We show this claim in the proof of Lemma \ref{lem:policyobjective}.

M$^{\natural}$-convexity is a weakening of the following property:
A set of distributions $\Xi\subseteq \Xi^0$ is \textbf{M-convex} if, for every $\xi,\tilde{\xi} \in \Xi$ and $(c,t) \in \calc \times \calt$ with $\xi_c^t>\tilde{\xi}_c^t$,
there exists $(c',t') \in \calc \times \calt$ with $\xi_{c'}^{t'}<\tilde{\xi}_{c'}^{t'}$ such that
$\xi-\chi_c^t+\chi_{c'}^{t'}\in \Xi \; \mbox{ and } \; \tilde{\xi}+\chi_c^t-\chi_{c'}^{t'} \in \Xi$.

To see the difference between these two concepts, consider the following simple examples:
$\{(2,0),(1,1),(0,2)\}$ is M-convex, and hence M$^{\natural}$-convex, whereas $\{0,1,2\}$ is M$^{\natural}$-convex but not M-convex.

\subsection{A Characterization of Pseudo M$^{\natural}$-concavity}\hfill\\
Here, we establish a close
connection between pseudo M$^{\natural}$-concavity of a function and M$^{\natural}$-convexity of a policy goal.


 Given a distributional objective $f$ and a constant $\lambda \in \mathbb{R}$, consider the following distributional policy goal, named $\boldsymbol{(f,\lambda)}$\textbf{-goal}:
$\Xi(f,\lambda) \equiv \{\xi \in \Xi^0 \mid f(\xi)\geq \lambda\}$. In words, a distribution satisfies the $(f,\lambda)$-goal if the value that $f$ takes
on the distribution is at least $\lambda$. 
In other words, $(f,\lambda)$-goals represent upper contour sets of function $f$.

We show that pseudo M$^{\natural}$-concavity is characterized by all upper contour sets being M$^{\natural}$-convex.
Moreover, for each M$^{\natural}$-convex policy goal, there exists a corresponding distributional objective satisfying pseudo M$^{\natural}$-concavity.

\begin{theorem}\label{thm:characterization}
The policy goal $\Xi(f,\lambda)$ is M$^{\natural}$-convex for every $\lambda \in \mathbb{R}$ if and only if the distributional objective
$f$ is pseudo M$^{\natural}$-concave. Moreover, if $\Xi \subseteq \Xi^0$ is M$^{\natural}$-convex, then there exist a constant $\lambda \in \mathbb{R}$ and
a pseudo M$^{\natural}$-concave distributional objective $f$ such that $\Xi(f,\lambda)  = \Xi$.
\end{theorem}

Theorem \ref{thm:characterization} demonstrates that pseudo M$^{\natural}$-concavity is a natural discrete counterpart of  the familiar notion of \emph{quasi-concavity}: A real-valued function $g$ defined on
a convex domain $D\subseteq \mathbb{R}^n$ is quasi-concave if, for every real number $\lambda\in \mathbb{R}$, the upper contour set $\{x\in D : g(x)\geq \lambda\}$ is
convex.

We also note that, similar to Theorem \ref{thm:characterization}, it can be shown that pseudo M-concavity is equivalent to the requirement that all upper contour sets are M-convex.

\subsection{Distributional Objectives Based on Policy Goals.}\hfill\\
Our previous analysis of distributional policy objectives has implications for desirable mechanisms that satisfy a policy goal.
To make this connection, we consider a diversity objective that is inversely related to the distance between the distribution
in consideration and the policy goal.
For this purpose, we use two different notions of distance.\footnote{In Appendix \ref{app:distances}, we consider
a third distributional policy objective based on the Manhattan distance (or $L_1$ metric) to a policy goal.}

The first is the Chebyshev distance (also called the maximum or $L_\infty$ metric). 
For any $\xi,\tilde{\xi} \in \Xi^0$,
\begin{equation*}
D_C(\xi,\tilde{\xi}) \equiv \max_{(c,t) \in \calc \times \calt} \mathlarger{\mid} \xi_c^t-\tilde{\xi}_c^t \mathlarger{\mid},\footnote{For a real number $x\in \mathbb R$, $\abs{x}$ denotes the absolute value of $x$.}
\end{equation*}
is the distance between two distributions $\xi$ and $\tilde{\xi}$. Then the distance between a distribution $\xi$
and a policy goal $\Xi$ is defined as $\min_{\tilde{\xi} \in \Xi} D_C(\xi,\tilde{\xi})$.
Hence, we consider the distributional objective $f_C^{\Xi}$, which is equal to the negative of the Chebyshev distance
between the policy goal $\Xi$ and the distribution in question.\footnote{Since higher values of $f$
means the distributional policy objective is satisfied more, we consider the negative of the distance to define the distributional objective.}
More explicitly, $f_C^{\Xi}$ is defined as, for every $\xi\in \Xi^0$,
\begin{equation*}
f_C^{\Xi}(\xi) \equiv -\min_{\tilde{\xi} \in \Xi} D_C(\xi,\tilde{\xi}). 
\end{equation*}

The next lemma shows that M$^{\natural}$-convexity of $\Xi$ is both necessary and sufficient for $f^{\Xi}_C$ to be pseudo M$^{\natural}$-concave.

\begin{lemma}\label{lem:cheb}
The policy goal $\Xi \subseteq \Xi^0$ is M$^{\natural}$-convex if and only if $f^{\Xi}_C$ is pseudo M$^{\natural}$-concave.
\end{lemma}

Next, we investigate an alternative distributional objective based on policy goals. Policymakers may be strict in implementing a policy goal and view a distribution as a success
if it is in the policy goal and everything outside of the policy goal as a failure. Therefore, the distributional objective may be defined using the discrete metric
(or the characteristic function) as follows. For each policy goal $\Xi$, the distributional objective $f^{\Xi}_D$ is given by
$f^{\Xi}_D(\xi)=1$ if $\xi \in \Xi$, and $f^{\Xi}_D(\xi)=0$, when $\xi \in \Xi^0 \setminus \Xi$.

The next lemma shows that M$^{\natural}$-convexity of $\Xi$ is both necessary and sufficient for $f^{\Xi}_D$ to be pseudo M$^{\natural}$-concave.

\begin{lemma}\label{lem:policyobjective}
The policy goal $\Xi \subseteq \Xi^0$ is M$^{\natural}$-convex if and only if $f^{\Xi}_D$ is pseudo M$^{\natural}$-concave.
\end{lemma}

The following result follows from Lemma \ref{lem:policyobjective} and Theorem \ref{thm:ttc}.

\begin{corollary}\label{cor:ttcp}
Suppose that $\mu_0$ satisfies the policy goal $\Xi \subseteq \Xi^0$. If $\Xi$ is M$^{\natural}$-convex, then the TTC mechanism for the distributional
objective $f^{\Xi}_D$
satisfies the policy goal $\Xi$, constrained efficiency, individual rationality, and strategy-proofness.
\end{corollary}

Since $\mu_0$ satisfies the policy goal $\Xi$, we have $f^{\Xi}_D(\xi(\mu_0))=1$. In addition, by Lemma \ref{lem:policyobjective},  $f^{\Xi}_D$
is pseudo M$^{\natural}$-concave. Therefore, by Theorem \ref{thm:ttc}, the TTC mechanism weakly improves
 $f^{\Xi}_D$. Since the value of  $f^{\Xi}_D$ is one if the distribution satisfies the policy goal $\Xi$ and zero otherwise, we conclude that
the TTC mechanism satisfies the policy goal $\Xi$. The other properties also immediately follow from Theorem \ref{thm:ttc}.

Corollary \ref{cor:ttcp} is a generalization of the main result of \cite{suzuki17}, who study a setting where, (i) there are no student types,\footnote{\cite{suzuki17}
state that type-specific floors and ceilings at schools form M-convex distributional constraints. \cite{suzuki2023strategyproof} show that the transformation
used in \cite{interdistrictwp}, which our paper is based on, can be applied for this specific application so that the main result in \cite{suzuki17} applies. However, the model of \cite{suzuki17} and \cite{suzuki2023strategyproof} do not have student types.}
(ii) students are not allowed to be unmatched, and (iii) the policy goal is M-convex. In Corollary \ref{cor:ttcp}, however, there can be multiple student types,
not all students need to be initially or eventually matched, and the policy goal is M$^{\natural}$-convex (rather than M-convex, which is more restrictive).

Note that the assumption that the initial matching satisfies the policy goal is necessary for Corollary \ref{cor:ttcp}. To see this, consider student preferences such that each student's highest-ranked option is her initial assignment. Then the initial matching is the unique individually rational matching. Therefore, if there exists a mechanism with the desired properties, then the outcome of this preference profile has to be the initial matching. Hence, we need the assumption that the initial matching satisfies the policy goal to have such a mechanism.

\section{Applications}\label{sec:applications}
In this section, we consider several real-life applications and show that a wide variety of
distributional policies can be expressed by distributional objectives that are pseudo M$^{\natural}$-concave
and, therefore, our TTC mechanism can be applied in these settings. Additional examples are provided in Appendix \ref{app:additional}.

\subsection{School-level quota policies}\label{sec:exchange}\hfill\\
In our first application, each school has a floor and ceiling on the number of students that it can admit. This is motivated by exchange markets, where participating institutions
either require the numbers of incoming and outgoing agents to be the same or set bounds on the imbalance between the incoming and outgoing agents \citep{dur2015two}.
More specifically, consider the following policy:
\[\Xi^Q \equiv \left \{\xi \in \Xi^0 \mathlarger{\mathlarger{\mid}}  u_c \geq \sum_{t} \xi_c^t \geq l_c \text{ for all } c\in \calc \right \},\]
where $u_c\in \mathbb N$ is the ceiling and $l_c \in \mathbb N$ is the floor at school $c\in \calc$ such that $u_c \geq l_c$.\footnote{\cite{dur2015two} make the
stronger assumption $u_c \geq \sum_{t} \xi_c^t(\mu_0) \geq l_c$, which we do not need.}
We call a policy goal $\Xi^Q$ of this form a \df{school-level quota policy}.

In Appendix \ref{sec:proofs}, we show that $\Xi^Q$ is an M$^{\natural}$-convex set. Therefore, we get the following
result using Theorem \ref{thm:ttc} and Lemmas \ref{lem:cheb} and \ref{lem:policyobjective}.

\begin{proposition}\label{prop:quotas}
Suppose that the distributional objective is $f_C^{\Xi^Q}$ or $f_D^{\Xi^Q}$. Then the TTC mechanism weakly improves the distributional objective and satisfies constrained efficiency, individual rationality, and strategy-proofness.
\end{proposition}

\subsection{School-level diversity policies}\label{sec:schoollevel}\hfill\\
In this application, we study diversity (or affirmative action) policies that are commonly used in controlled school choice \citep{ehayeyi14}.
Suppose that the policy goal sets type-specific floors and ceilings at each school. More precisely,
\[\Xi^D \equiv \left \{\xi \in \Xi^0 \mathlarger{\mathlarger{\mid}} q_c^t \geq \xi_c^t \geq p_c^t \text{ for all } (c,t)\in \calc \times \calt\right \}\]
where $q_c^t\in \mathbb N$ is the ceiling and $p_c^t\in \mathbb N$ is the floor for type $t\in \calt$ at school $c\in \calc$ such that $q_c^t  \geq p_c^t$.
This policy requires, for each school, the number of students of a given type to be between the ceiling and floor of this type at the school.
As it is standard in the controlled school choice literature, we assume that each school $c\in \calc$ has enough capacity to cover all floors,
that is, $q_c \geq \sum_t p_c^t$. We call a policy goal of this form a \df{school-level diversity policy}.

In Appendix \ref{sec:proofs}, we show that $\Xi^D$ is an M$^{\natural}$-convex set.
Therefore, Theorem \ref{thm:ttc} and Lemmas \ref{lem:cheb} and \ref{lem:policyobjective} imply the following result.

\begin{proposition}\label{prop:convexdiv}
Suppose that the distributional objective is $f_C^{\Xi^D}$ or $f_D^{\Xi^D}$. Then the TTC mechanism weakly improves the distributional objective and satisfies constrained efficiency, individual rationality, and strategy-proofness.
\end{proposition}

\subsection{Exchange-feasibility policy}\label{subsec:bf}\hfill\\
Next, we introduce a variation of the balanced-exchange policy studied in \cite{hafalir2022interdistrict}.
The motivation comes from a school choice problem with school districts
where students can be assigned to schools outside of their home districts. Each school district would like to have the number of students assigned to schools in
the district to not decrease compared to the initial matching. This is a practically important objective because, for example, in the U.S., each school district's funding increases in the number of students that it enrolls.\footnote{For example, a 2005 report details the state funding for school districts in the U.S., which depends on the number of
enrolled students, in all states except Delaware, Hawaii, and Pennslyvania. See \href{https://www.ecs.org/clearinghouse/59/81/5981.pdf}{https://www.ecs.org/clearinghouse/59/81/5981.pdf} (last accessed on January 4, 2023).}

Suppose that there is a set of districts $\cald$ and each school $c\in \calc$ is in a district $d(c)\in \cald$.
With the exchange-feasibility policy, for each district, we require the total number of
students assigned to schools in the district to weakly increase. Let $k_{d}$ denote
the number of students assigned to schools in district $d\in \cald$ at
the initial matching.\footnote{Propositions \ref{prop:convexbal} and \ref{prop:mix} below hold for every $(k_d)_{d\in \cald}$ because M$^{\natural}$-convexity
of $\Xi^E$ and $\Xi^{DE}$ do not depend on the profile $(k_d)_{d\in \cald}$.} Then, the \df{exchange-feasibility policy} is formally defined as
\[\Xi^E \equiv \left \{\xi \in \Xi^0 \mathlarger{\mathlarger{\mid}} \sum_{t,c:d(c)=d} \xi_c^t \geq k_{d} \text{ for all } d \in \cald \right \}.\]
As before, we show that $\Xi^E$ is an M$^{\natural}$-convex set in Appendix \ref{sec:proofs} and, therefore, the following result holds.
\begin{proposition}\label{prop:convexbal}
Suppose that the distributional objective is $f_C^{\Xi^E}$ or $f_D^{\Xi^E}$. Then the TTC mechanism weakly improves the distributional objective and satisfies constrained efficiency, individual rationality, and strategy-proofness.
\end{proposition}

One of the advantages of our approach is that pseudo M$^ {\natural}$-concavity of a distributional objective (or M$^{\natural}$-convexity of a policy goal)
is so general that a wide variety of distributional objectives satisfy it and, therefore, it is likely to be applicable for distributional objectives that one may
encounter in the future. To highlight this point, we consider school-level diversity 
and exchange-feasibility policies simultaneously. More specifically, let
\[\Xi^{DE} \equiv \left\{\xi \in \Xi^0 \mathlarger{\mathlarger{\mid}} q_c^t \geq \xi_c^t \geq p_c^t \mbox{ , } \forall (c,t)\in \calc \times \calt \mbox{ and }  \sum_{t,c:d(c) = d} \xi_c^t \geq k_{d} \text{ , } \forall d \in \cald \right\}\]
and call it the \df{combination of school-level diversity and exchange-feasibility policies}.
This is the set of distributions that satisfy the school-level diversity policies and the flow-feasibility requirement for districts. We
establish that $\Xi^{DE}$ is M$^{\natural}$-convex, implying the following result.

\begin{proposition}\label{prop:mix}
Suppose that the distributional objective is $f_C^{\Xi^{DE}}$ or $f_D^{\Xi^{DE}}$. Then the TTC mechanism weakly improves the distributional objective and satisfies constrained efficiency, individual rationality, and strategy-proofness.
\end{proposition}

In general, the intersection of two M$^{\natural}$-convex sets need not be M$^{\natural}$-convex.\footnote{Such
an example is available from the authors upon request.} Therefore, the proof of this result
does not follow from the proofs of Propositions \ref{prop:convexdiv} and \ref{prop:convexbal}.

\subsection{District-level Diversity Policies}\hfill\\
We consider the interdistrict school choice setting as in Section \ref{subsec:bf}, but instead of the
exchange-feasibility policy, we study diversity policies at the district level as follows. Suppose that,
for each district $d \in \cald$ and type $t \in \calt$, there exist a type-specific floor $p_d^t\in \mathbb N$
and a type-specific ceiling $q_d^t\in \mathbb N$ at the district level such that $q_d^t \geq p_d^t$. Therefore, the policy goal can be written as
\[\left \{\xi \in \Xi^0 \mathlarger{\mathlarger{\mid}} q_d^t \geq  \sum_{c:d(c)=d} \xi_c^t \geq p_d^t \text{ for all } (d,t)\in \cald \times \calt \right \}.\]
Moreover, suppose that the floors and ceilings are not too tight in the sense that there exist feasible distributions that satisfy the policy goal.
We show that this policy goal is not necessarily M$^{\natural}$-convex and, therefore, the distributional objectives
corresponding to this policy goal that we defined in Section \ref{sec:alternative} are not pseudo M$^{\natural}$-concave
(Lemmas \ref{lem:cheb} and \ref{lem:policyobjective}).

To see this, consider the setting in Example \ref{ex:notconvex}. Suppose that there are two districts: $d_1$ has schools $c_1$, $c_2$, and $c_3$,
whereas $d_2$ has schools $c_4$, $c_5$, and $c_6$. Let the floors and ceilings be $q_d^t = p_d^t = 1$ for each type $t\in \calt$ and district $d\in \cald$.
We have shown in Section \ref{sec:TTC} that the distributional objective corresponding to this policy goal for the discrete metric
fails pseudo M$^{\natural}$-concavity. Therefore, the policy goal is not M$^{\natural}$-convex (Lemma \ref{lem:policyobjective}).
Hence, the distributional objective corresponding to this policy goal using the Chebyshev distance also fails pseudo M$^{\natural}$-concavity (Lemma \ref{lem:cheb}).

We note that the main reason for the failure of M$^{\natural}$-convexity is not floors (since floors
do not create a problem as we have seen in Section \ref{subsec:bf}). This is in contrast with the controlled school choice
literature in the context of 
\emph{stable} matchings: \cite{abdulson03} establish
the existence of stable matchings when there are type-specific ceilings at each school whereas \citet{ehayeyi14}
show the failure of the stable matchings when there are type-specific floors at schools.

\section{Conclusion}\label{sec:conclusion}
We studied the principles that Keynes laid down as the political problem of mankind in a matching context. In our setting,
the challenge was to design a mechanism that achieves constrained efficiency, individual rationality, and strategy-proofness
while weakly improving a distributional objective. We identified a notion of discrete concavity, which we called
pseudo M$^{\natural}$-concavity, and showed that when the distributional objective satisfies this notion, there exists a mechanism
with all the desirable properties listed above.  In fact, we provided an explicit mechanism that achieves these goals, namely a
generalization of the celebrated top-trading cycles mechanism \citep{shasca74}. 

One of our main contributions is to identify a general class of distributional objectives, rather than a specific objective,
under which a mechanism with the desirable properties exists. Pseudo M$^{\natural}$-concavity is quite permissive in the sense
of being satisfied by many important real-life applications. Identifying additional practical examples of distributional
objectives that are pseudo M$^{\natural}$-concave is left for future research.

We hope that our approach will pave the way for a more systematic analysis of distributional policies. For example, in our
context, whether pseudo M$^{\natural}$-concavity can be weakened for the existence of a mechanism with the desirable properties
is an important open question and is also left for future research. We hope and anticipate that our results
will be useful in attaining a better understanding of when constrained efficiency can be achieved in matching environments where distributional policies play a crucial role.

\bibliographystyle{aer}
\bibliography{matching}


\appendix

\section{An Illustration of the TTC Algorithm}\label{subsec:ttcexample}
In this appendix, we illustrate how the TTC algorithm works. Consider a matching problem
with $\cals=\{s_1,\ldots,s_7\}$, $\calc=\{c_1, c_2, c_3, c_4\}$, and $\calt=\{t_1,t_2\}$.
School $c_{1}$ has capacity three, school $c_{2}$ has capacity two, and schools $c_{3}$ and $c_{4}$ have capacity one.
Students $s_{1}$, $s_{2}$, $s_{3}$, and $s_{4}$ have type $t_{1}$ whereas students $s_{5}$, $s_{6}$, and $s_{7}$
have type $t_{2}$. The initial matching $\mu_0$ is $\{(s_1,c_1),(s_2,c_1),(s_3,c_2),(s_4,\emptyset),(s_5,\emptyset),(s_6,c_3),(s_7,c_4)\}$.
Student preferences are as follows.

\[
\begin{tabular}
[c]{lllllll}
$\underline{P_{s_{1}}}$ & $\underline{P_{s_{2}}}$ & $\underline{P_{s_{3}}}$ &
$\underline{P_{s_{4}}}$ & $\underline{P_{s_{5}}}$ & $\underline{P_{s_{6}}}$ &
$\underline{P_{s_{7}}}$\\
$c_{2}$ & $c_{3}$ & $c_{4}$ & $c_{3}$ & $c_{1}$ & $c_{4}$ & $c_{2}$\\
$c_3 $ & $c_{1}$ & $c_{2}$ & $c_{1}$ & $c_{2}$ & $c_3$ & $c_{3}$\\
 $c_1$ & $\vdots$ & $\vdots$ & $\emptyset$ & $\emptyset$ & $\vdots$ & $c_{4}$\\
$\vdots$ &  &  & $\vdots$ & $\vdots$ & & $\vdots$%
\end{tabular}
\]
\smallskip

The distributional policy objective is given by the following: (i) $f(\xi) = 1$ if $\xi$ is a feasible distribution,
$\xi_{c_1}^{t_1} \leq 2$, $\xi_{c_1}^{t_2} \leq 1$, $\xi_{c_2}^{t_1} \leq 1$, and $\xi_{c_2}^{t_2} \leq 1$;  (ii) $f(\xi) = 0$, otherwise. It is easy to verify that $f(\xi(\mu_0)) = 1$.

To run the TTC algorithm, we use a master priority list. Suppose that the master priority list ranks students as follows: $s_1 \succ s_2 \succ s_3 \succ s_4 \succ s_5 \succ s_6 \succ s_7$.

\begin{figure}[!htb]
\begin{center}
    \begin{tikzpicture}[scale = .55, font = {\footnotesize}]
      \coordinate (O)    at (2, 0);
      \coordinate (s1)   at (2, 6);
      \coordinate (s2)   at (2, 2);
      \coordinate (s3)   at (2, -2);
      \coordinate (s4)   at (0, 0);
      \coordinate (s5)   at (-1, 2);
      \coordinate (s6)   at (-4, 4);
      \coordinate (s7)   at (-2, 6);
      \coordinate (c1t1) at (0, 6);
      \coordinate (c1t2) at (0, 4);
      \coordinate (c2t1) at (4, 4);
      \coordinate (c2t2) at (5, 1);
      \coordinate (c3t1) at (-2, 0);
      \coordinate (c3t2) at (-4, 2);
      \coordinate (c4t1) at (-6, 4);
      \coordinate (c4t2) at (-4, 6);

      \node (0)    [school s, label = {[shift = {(0.15, -0.1)}] below:$(\emptyset, \emptyset)$}] at (O) {};
      \node (s1)   [school n = $s_1$] at (s1) {};
      \node (s2)   [school n = $s_2$] at (s2) {};
      \node (s3)   [school s = $s_3$] at (s3) {};
      \node (s4)   [school s = $s_4$] at (s4) {};
      \node (s5)   [school w = $s_5$] at (s5) {};
      \node (s6)   [school w = $s_6$] at (s6) {};
      \node (s7)   [school s = $s_7$] at (s7) {};
      \node (c1t1) [school n, label = {[shift = {(0, 0)}] above:$(c_1, t_1)$}] at (c1t1) {};
      \node (c1t2) [school w, label = {[shift = {(-.2, .3)}] left:$(c_1, t_2)$}] at (c1t2) {};
      \node (c2t1) [school e, label = {[shift = {(.4, 0)}] right:$(c_2, t_1)$}] at (c2t1) {};
      \node (c2t2) [school e, label = {[shift = {(.4, 0)}] right:$(c_2, t_2)$}] at (c2t2) {};
      \node (c3t1) [school w, label = {[shift = {(-.4, 0)}] left:$(c_3, t_1)$}] at (c3t1) {};
      \node (c3t2) [school w, label = {[shift = {(0, 0)}] below:$(c_3, t_2)$}]  at (c3t2) {};
      \node (c4t1) [school w, label = {[shift = {(-.4, 0)}] left:$(c_4, t_1)$}] at (c4t1) {};
      \node (c4t2) [school w, label = {[shift = {(-.27, 0)}] left:$(c_4, t_2)$}] at (c4t2) {};

      \fill (2.2,0)    circle (2pt);
      \fill (s1)   circle (2pt);
      \fill (s2)   circle (2pt);
      \fill (s3)   circle (2pt);
      \fill (s4)   circle (2pt);
      \fill (s5)   circle (2pt);
      \fill (s6)   circle (2pt);
      \fill (s7)   circle (2pt);
      \fill (c1t1) circle (2pt);
      \fill (c1t2) circle (2pt);
      \fill (c2t1) circle (2pt);
      \fill (c2t2) circle (2pt);
      \fill (c3t1) circle (2pt);
      \fill (c3t2) circle (2pt);
      \fill (c4t1) circle (2pt);
      \fill (c4t2) circle (2pt);

      \draw [arrow]   (O)    to                                   (s4);
      \draw [arrow]   (s1)   to                                   (c2t1);
      \draw [arrow]   (s2)   to [bend right = 15]                 (c3t1);
      \draw [t arrow] (s3)   to [bend left = 60]                  (c4t1);
      \draw [arrow]   (s4)   to                                   (c3t1);
      \draw [arrow]   (s5)   to                                   (c1t2);
      \draw [arrow]   (s6)   to                                   (c4t2);
      \draw [t arrow] (s7)   to [bend left = 90, looseness = 2]   (c2t2);
      \draw [arrow]   (c1t1) to                                   (s1);
      \draw [arrow]   (c1t2) to                                   (s1);
      \draw [arrow]   (c2t1) to [bend left = 30]                  (s3);
      \draw [t arrow] (c2t2) to [bend left = 30]                  (s3);
      \draw [arrow]   (c3t1) to                                   (s6);
      \draw [arrow]   (c3t2) to                                   (s6);
      \draw [t arrow] (c4t1) to [bend left = 100, looseness = 1.7] (s7);
      \draw [arrow]   (c4t2) to                                   (s7);
    \end{tikzpicture}
\end{center}
  \caption{First step of the TTC algorithm. The cycle is represented by thick arrows.}
  \label{fig:TTCstep1}
\end{figure}
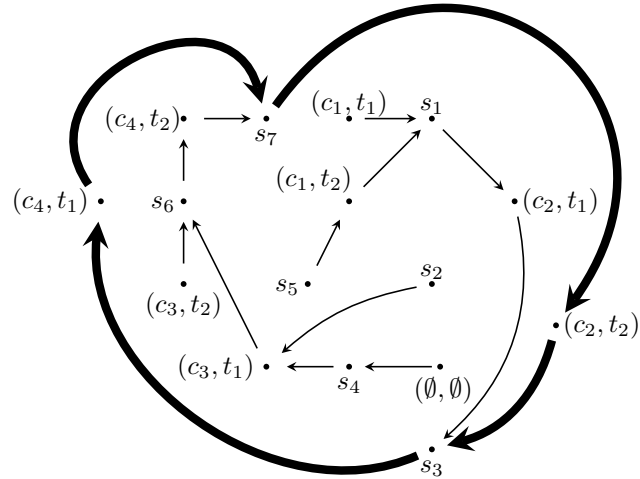

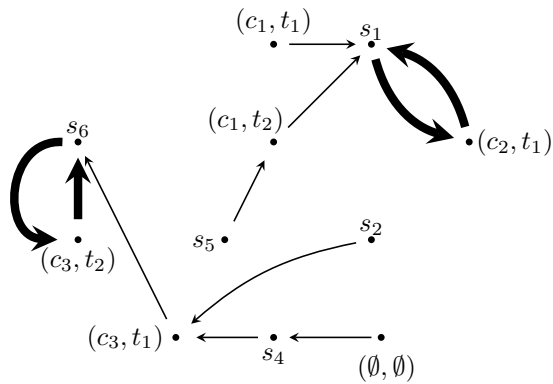
\begin{figure}
\begin{center}
    \begin{tikzpicture}[scale = .65, font = {\footnotesize}]
      \coordinate (O)    at (2, 0);
      \coordinate (s1)   at (2, 6);
      \coordinate (s2)   at (2, 2);
      \coordinate (s3)   at (2, -2);
      \coordinate (s4)   at (0, 0);
      \coordinate (s5)   at (-1, 2);
      \coordinate (s6)   at (-4, 4);
      \coordinate (s7)   at (-2, 6);
      \coordinate (c1t1) at (0, 6);
      \coordinate (c1t2) at (0, 4);
      \coordinate (c2t1) at (4, 4);
      \coordinate (c2t2) at (5, 1);
      \coordinate (c3t1) at (-2, 0);
      \coordinate (c3t2) at (-4, 2);
      \coordinate (c4t1) at (-6, 4);
      \coordinate (c4t2) at (-4, 6);

      \node (0)    [school s, label = {[shift = {(0.2, -0.2)}] below:$(\emptyset, \emptyset)$}] at (O) {};
      \node (s1)   [school n = $s_1$] at (s1) {};
      \node (s2)   [school n = $s_2$] at (s2) {};
      \node (s4)   [school s = $s_4$] at (s4) {};
      \node (s5)   [school w = $s_5$] at (s5) {};
      \node (s6)   [school n = $s_6$] at (s6) {};
      \node (c1t1) [school n, label = {[shift = {(0, .1)}]    above:$(c_1, t_1)$}] at (c1t1) {};
      \node (c1t2) [school w, label = {[shift = {(-.1, .30)}] left: $(c_1, t_2)$}] at (c1t2) {};
      \node (c2t1) [school e, label = {[shift = {(.4, 0)}]   right:$(c_2, t_1)$}] at (c2t1) {};
      \node (c3t1) [school w, label = {[shift = {(-.4, 0)}]  left: $(c_3, t_1)$}] at (c3t1) {};
      \node (c3t2) [school w, label = {[shift = {(0, 0)}]    below:$(c_3, t_2)$}] at (c3t2) {};

      \fill (2.2,0)    circle (2pt);
      \fill (s1)   circle (2pt);
      \fill (s2)   circle (2pt);
      \fill (s4)   circle (2pt);
      \fill (s5)   circle (2pt);
      \fill (s6)   circle (2pt);
      \fill (c1t1) circle (2pt);
      \fill (c1t2) circle (2pt);
      \fill (c2t1) circle (2pt);
      \fill (c3t1) circle (2pt);
      \fill (c3t2) circle (2pt);

      \draw [arrow]   (O)    to (s4);
      \draw [t arrow] (s1)   to [bend right = 30] (c2t1);
      \draw [arrow]   (s2)   to [bend right = 15] (c3t1);
      \draw [arrow]   (s4)   to (c3t1);
      \draw [arrow]   (s5)   to (c1t2);
      \draw [t arrow] (s6)   to [bend right = 90, looseness = 1.5] (c3t2);
      \draw [arrow]   (c1t1) to (s1);
      \draw [arrow]   (c1t2) to (s1);
      \draw [t arrow] (c2t1) to [bend right = 30] (s1);
      \draw [arrow]   (c3t1) to (s6);
      \draw [t arrow] (c3t2) to (s6);
    \end{tikzpicture}
\end{center}
  \caption{Second step of the TTC algorithm. Cycles are represented by thick arrows.}
  \label{fig:TTCstep2}
\end{figure}

At Step $1$, there are eight school-type pairs and $(\emptyset,\emptyset)$ which represents being unassigned. Consider $(c_1,t_1)$. Initially, students $s_1$ and $s_2$ are matched with $c_1$, so they are both permissible to this pair. We use the master priority list to rank them, so $s_1$ gets the highest priority at $(c_1,t_1)$. Therefore, $(c_1,t_1)$ points to $s_1$. Now consider $(c_1,t_2)$. Initially, it does not have any students because there is no type-$t_2$ student assigned to $c_1$ in the original matching problem. Furthermore, $s_1$ is permissible to
$(c_1,t_2)$ because she can be removed from $(c_1,t_1)$ and a type-$t_2$ student can be assigned to $(c_1,t_2)$ without worsening the distributional policy objective. Therefore, $(c_1,t_2)$ points to $s_1$ as well because $s_1$ gets a higher priority than other permissible students because of the master priority list. The rest of the pairs also point to the highest-priority permissible students. Each student points to the highest ranked school-type pair of the same type as shown in Figure \ref{fig:TTCstep1}. There is only one cycle at Step $1$: $s_7 \rightarrow (c_2,t_2) \rightarrow s_3 \rightarrow (c_4,t_1) \rightarrow s_7$. Therefore, $s_7$ is assigned to $(c_2,t_2)$ and $s_3$ is assigned to $(c_4,t_1)$.
At Step $2$, there are five remaining school-type pairs and $(\emptyset,\emptyset)$: there are no permissible students for $(c_4,t_1)$ and $(c_4,t_2)$ because $c_4$ has a
capacity of one and it is already assigned to $s_3$; there are no permissible students for $(c_2,t_2)$ because $c_2$ is already assigned to a type-$t_2$ student.
Each remaining school-type pair points to the highest-ranked remaining permissible student. Each student points to the highest-ranked remaining school-type pair
(see Figure \ref{fig:TTCstep2}). There are two cycles: $s_1 \rightarrow (c_2,t_1) \rightarrow s_1$ and $s_6 \rightarrow (c_3,t_2) \rightarrow s_6$.
Hence, $s_1$ is assigned to $(c_2,t_1)$ and $s_6$ is assigned to $(c_3,t_2)$.

\begin{figure}[!htb]
\centering
  \subfloat[Step 3 of TTC]{
    \label{fig:TTCstep3}
    \begin{tikzpicture}[scale = .80, font = {\footnotesize}]
      \coordinate (O)    at (2, 0);
      \coordinate (s1)   at (2, 6);
      \coordinate (s2)   at (2, 2);
      \coordinate (s3)   at (2, -2);
      \coordinate (s4)   at (0, 0);
      \coordinate (s5)   at (-1, 2);
      \coordinate (s6)   at (-4, 4);
      \coordinate (s7)   at (-2, 6);
      \coordinate (c1t1) at (4, 3);
      \coordinate (c1t2) at (0, 4);
      \coordinate (c2t1) at (4, 4);
      \coordinate (c2t2) at (5, 1);
      \coordinate (c3t1) at (-2, 0);
      \coordinate (c3t2) at (-4, 2);
      \coordinate (c4t1) at (-6, 4);
      \coordinate (c4t2) at (-4, 6);

      \node (0)    [school s, label = {[shift = {(0.2, -0.2)}] below:$(\emptyset, \emptyset)$}] at (O) {};
      \node (s2)   [school s = $s_2$] at (s2) {};
      \node (s4)   [school s = $s_4$] at (s4) {};
      \node (s5)   [school w = $s_5$] at (s5) {};
      \node (c1t1) [school n, label = {[shift = {(0.0, .2)}] above:$(c_1, t_1)$}] at (c1t1) {};
      \node (c1t2) [school w, label = {[shift = {(0.3, .4)}] left:$(c_1, t_2)$}] at (c1t2) {};

      \fill (2.2,0)    circle (2pt);
      \fill (s2)   circle (2pt);
      \fill (s4)   circle (2pt);
      \fill (s5)   circle (2pt);
      \fill (c1t1) circle (2pt);
      \fill (c1t2) circle (2pt);

      \draw [arrow]   (O)    to [bend left = 20] (s4);
      \draw [t arrow] (s2)   to [bend right = 20] (c1t1);
      \draw [arrow]   (s4)   to [bend right = 40] (c1t1);
      \draw [arrow]   (s5)   to (c1t2);
      \draw [t arrow] (c1t1) to [bend right = 20] (s2);
      \draw [arrow]   (c1t2) to (s2);
    \end{tikzpicture}
  }
  \hfill
  \subfloat[Step 4 of TTC]{
    \label{fig:TTCstep4}
    \begin{tikzpicture}[scale = .80, font = {\footnotesize}]
      \coordinate (O)    at (2, 0);
      \coordinate (s1)   at (2, 6);
      \coordinate (s2)   at (2, 2);
      \coordinate (s3)   at (2, -2);
      \coordinate (s4)   at (0, 0);
      \coordinate (s5)   at (-2, 4);
      \coordinate (s6)   at (-4, 4);
      \coordinate (s7)   at (-2, 6);
      \coordinate (c1t1) at (3, 3);
      \coordinate (c1t2) at (0, 4);
      \coordinate (c2t1) at (4, 4);
      \coordinate (c2t2) at (5, 1);
      \coordinate (c3t1) at (-2, 0);
      \coordinate (c3t2) at (-4, 2);
      \coordinate (c4t1) at (-6, 4);
      \coordinate (c4t2) at (-4, 6);

      \node (0)    [school s, label = {[shift = {(0.2, -0.2)}] below:$(\emptyset, \emptyset)$}] at (O) {};
      \node (s4)   [school s = $s_4$] at (s4) {};
      \node (s5)   [school w = $s_5$] at (s5) {};
      \node (c1t1) [school n, label = {[shift = {(0, 0.1)}] above:$(c_1, t_1)$}] at (c1t1) {};
      \node (c1t2) [school w, label = {[shift = {(.4, 0)}] right:$(c_1, t_2)$}] at (c1t2) {};

      \fill (2.2,0)    circle (2pt);
      \fill (s4)   circle (2pt);
      \fill (s5)   circle (2pt);
      \fill (c1t1) circle (2pt);
      \fill (c1t2) circle (2pt);

      \draw [arrow]   (O)    to [bend left = 10] (s4);
      \draw [t arrow] (s4)   to (c1t1);
      \draw [arrow]   (s5)   to (c1t2);
      \draw [t arrow]   (c1t1) to [bend left = 30] (s4);
      \draw [arrow] (c1t2) to (s4);
    \end{tikzpicture}
  }
  \hfill
  \caption{Steps three and four of the TTC algorithm. Cycles are represented by thick arrows.}
  \label{fig:TTC}
\end{figure}
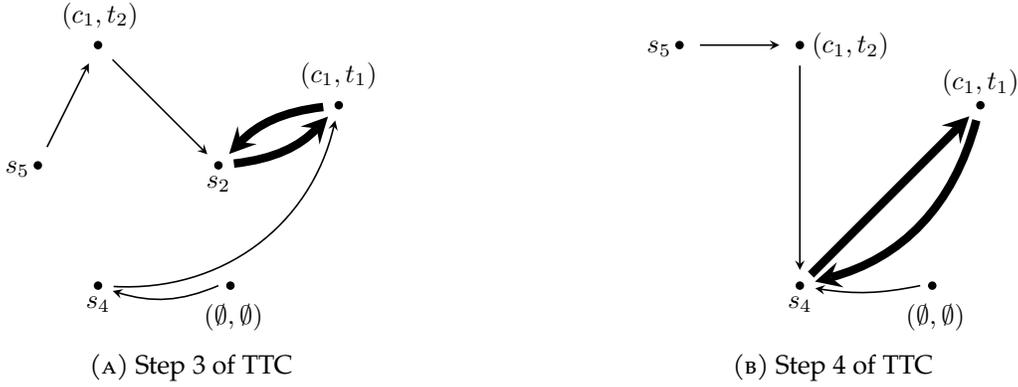

The algorithm ends in five steps. Steps 3 and 4 are shown in Figure \ref{fig:TTC}, Panels A and B. In  Step 5, $s_5$ points to $(c_1,t_2)$, which points back to $s_5$. The outcome is \[
\{(s_1,c_2),(s_2,c_1),(s_{3},c_{4}),(s_{4},c_{1}),(s_{5},c_1),(s_{6},c_{3}),(s_{7},c_{2})\}.
\]

It can be easily seen that the distribution induced by this matching weakly improves the distribution policy.

\section{Manhattan Distance}\label{app:distances}

In this appendix, we consider the Manhattan distance (or $L_1$ metric) between two
distributions and the corresponding distributional objective for policy sets. The Manhattan distance is defined as, for any $\xi, \tilde \xi \in \Xi^0$,
\begin{equation*}
D_M(\xi,\tilde{\xi}) \equiv  \sum_{(c,t)\in \calc \times \calt} \mathlarger{\mid} \xi_c^t-\tilde{\xi}_c^t \mathlarger{\mid}.
\end{equation*}
Using the Manhattan distance between two distributions, we define the Manhattan distance between a distribution $\xi$
and a policy set $\Xi$ as $\min_{\tilde{\xi} \in \Xi}  D_M(\xi,\tilde{\xi})$. Then, we consider the following
distributional objective:
\begin{equation*}
f_M^{\Xi}(\xi) \equiv - \min_{\tilde{\xi} \in \Xi}  D_M(\xi,\tilde{\xi}). 
\end{equation*}
We show that $f_M^{\Xi}$ is not necessarily pseudo M$^{\natural}$-concave even when $\Xi$ is M$^{\natural}$-convex.

\begin{example}
Consider a setting with one school $c$ and two types $t_1$ and $t_2$. Let $(x,y)$ represent the distribution where there are $x$
number of type-$t_1$ students and $y$ number of type-$t_2$ students. Suppose $\Xi=\{(1,1),(2,1)\}$.
It is easy to verify that $\Xi$ is M$^{\natural}$-convex.

Let $\xi=(3,1)$ and $\tilde{\xi}=(2,0)$ and note that $f_M^{\Xi}(\xi)=f_M^{\Xi}(\tilde{\xi})=-1$. Since $\xi_c^{t_1}>\tilde{\xi}_c^{t_1}$, pseudo M$^{\natural}$-concavity
implies the existence of $(c',t') \in \left( \calc \times \calt \right) \cup \{(\emptyset,\emptyset)\}$
(with $\xi_{c'}^{t'}<\tilde{\xi}_{c'}^{t'}$ whenever $(c',t')\neq (\emptyset,\emptyset)$) such that
\begin{equation*}
  \min \{f_M^{\Xi}(\xi-\chi_c^{t_1}+\chi_{c'}^{t'}), f_M^{\Xi}(\tilde{\xi}+\chi_c^{t_1}-\chi_{c'}^{t'})\} \geq \min \{f_M^{\Xi}(\xi),f_M^{\Xi}(\tilde{\xi})\}.
\end{equation*}
Since there is no coordinate in which $\tilde\xi$ is greater than $\xi$, it must be that $(c',t')=(\emptyset,\emptyset)$. Therefore,
$\xi-\chi_c^{t_1}+\chi_{c'}^{t'}=(2,1)$ and $\tilde{\xi}+\chi_c^{t_1}-\chi_{c'}^{t'}=(3,0)$. It is easy to see that
$f_M^{\Xi}(2,1)=0$ while $f_M^{\Xi}(3,0)=-2$. Therefore, the left-hand side of the displayed inequality is $-2$, whereas
the right-hand side of the displayed inequality is $-1$. We conclude that
$f_M^{\Xi}$ is not necessarily pseudo M$^{\natural}$-concave even though $\Xi$ is M$^{\natural}$-convex.
\qed
\end{example}

\section{Additional Results}\label{app:additional}

In this section, we provide additional markets with distributional objectives that are pseudo M$^{\natural}$-concave and establish
new results as applications of Theorem \ref{thm:ttc}.

First, we study the balanced-exchange policy requiring that the number of students assigned to schools in the district to remain the same compared to the initial matching.\footnote{This policy is equivalent to the balanced exchange policy considered by \cite{hafalir2022interdistrict} when all students are required to be matched.} As in Section \ref{subsec:bf}, we assume that there is a set of districts denoted by $\cald$ and
schools are partitioned into districts where the district of school $c$ is denoted by $d(c) \in \cald$.

With a balanced-exchange policy, we require the
number of students assigned to schools in each district to remain the same. More specifically, when the number of students assigned to schools in district $d$ at the initial matching is denoted by $k_{d}$,  the \textbf{balanced-exchange policy} can be written as
\[\Xi^{B} \equiv \left \{\xi \in \Xi^0 \mathlarger{\mathlarger{\mid}} \sum_{t,c:d(c)=d} \xi_c^t = k_{d} \text{ for all } d \in \cald \right \}.\]
The balanced-exchange policy is more restrictive than the exchange-feasibility policy $\Xi^E$ that we introduced in
Section \ref{subsec:bf}, so  $\Xi^{B} \subseteq \Xi^E$.

We show that $\Xi^{B}$ is M-convex (and, hence, M$^ {\natural}$-convex) in Appendix \ref{sec:proofs}.  Therefore, we get the following result.

\begin{proposition}\label{prop:bal-ex}
Suppose that the distributional objective is $f_C^{\Xi^{B}}$ or $f_D^{\Xi^B}$. Then the TTC mechanism weakly improves the distributional objective and satisfies constrained efficiency, individual rationality, and strategy-proofness.
\end{proposition}

Lastly, consider imposing school-level diversity and balanced-exchange policies for districts simultaneously.
More specifically, consider the following set of distributions
\[\Xi^{DB} \equiv \left \{\xi\in \Xi^0 \mid q_c^t \geq \xi_c^t \geq p_c^t \text{ , } \forall (c,t)\in \calc \times \calt \text{ and }
  \sum_{t,c:d(c) = d} \xi_c^t = k_{d} \text{ , } \forall d \in \cald\right \}\]
and call it the \df{combination of school-level diversity and balanced-exchange policies}. This is the set of distributions that satisfy both the school-level floors and ceilings and the balanced-exchange requirement. Since the balanced-exchange policy is more restrictive than the exchange-feasibility policy that we introduced
in Section \ref{subsec:bf}, we get $\Xi^{DB} \subseteq \Xi^{DE}$. Thus, the combination of school-level diversity and balanced-exchange policies is
more restrictive than the combination of school-level diversity and exchange-feasibility policies.

We establish that $\Xi^{DB}$ is M-convex (and, hence, M$^ {\natural}$-convex), implying the following result.

\begin{proposition}\label{prop:bal-mix}
Suppose that the distributional objective is $f_C^{\Xi^{DB}}$ or $f_D^{\Xi^{DB}}$. Then the TTC mechanism weakly improves the distributional objective and satisfies constrained efficiency, individual rationality, and strategy-proofness.
\end{proposition}

\section{Omitted Proofs}\label{sec:proofs}
In this section, we provide the omitted proofs where we use the following notation. For a distribution $\xi \in \mathbb N^{|\C|\times |\T|}$, let
$n(\xi) \equiv \sum_{c,t} \xi_c^t$
denote the number of students who are assigned at $\xi$, and,  for any $c \in \calc$,
$n_c(\xi) \equiv \sum_{t} \xi_c^t$
denote the number of students who are assigned to $c$ at $\xi$. When schools are partitioned into districts and, for each school $c$, its
district is denoted by $d(c)$,
$n_d(\xi) \equiv \sum_{t,c:d(c)=d} \xi_c^t$
denotes the number of students who are assigned to schools in district $d$ at a distribution
$\xi \in \mathbb N^{|\C|\times |\T|}$.

\begin{proof}[Proof of Theorem \ref{thm:ttc}]
We build on the main result of \cite{suzuki17}. They study a setting where each student is initially matched with a school and there is a policy goal $\Xi$. Furthermore, there are no student types and, hence, there are no constraints associated with student types.
In their setting, they show that if the policy goal $\Xi$ is M-convex and the initial distribution of students satisfies the policy
goal $\Xi$, then their mechanism, called TTC-M, satisfies the policy goal $\Xi$, constrained efficiency, individual rationality, and strategy-proofness.

Let us recap the hypothetical market that we introduced before the definition of the TTC algorithm in Section \ref{ttc_sec}. On one side of the market, there are school-type pairs $(c,t)$
where $c\in \C$ and $t\in \calt$, and also an outside option $(\emptyset,\emptyset)$ that represents being unmatched.
On the other side, there are students from the original problem, $\S$. The preferences of the students and
the priority orders of school-type pairs are given in Section \ref{ttc_sec}.

Next, let $\Xi \equiv \Xi(f,f(\mu_0))$ be the $(f,f(\mu_0))-$policy goal.
By Theorem \ref{thm:characterization}, the policy goal $\Xi$ is M$^{\natural}$-convex since $f$ is pseudo M$^{\natural}$-concave.

We now verify that the hypothetical market with a modified policy goal based on $\Xi$ satisfies all the conditions assumed by \cite{suzuki17}.
We denote the modified policy goal by $\tilde{\Xi}$. To construct it, we add a coordinate to $\Xi$
to represent the number of unassigned students. More formally,
$\tilde{\Xi} \equiv \{(\xi,|\mathcal S|- n(\xi)| \xi \in \Xi\}$, where $|\mathcal S|- n(\xi)$ is
the number of unassigned students in $\xi$.
In what follows, we argue that, when $\Xi$ satisfies M$^{\natural}$-convexity,  $\tilde{\Xi}$ satisfies M-convexity.
Theorem 6.3 of \cite{Murota:SIAM:2003}
and the discussion afterward establish that a M$^{\natural}$-convex set with $n$ variables results in a M-convex set with $n+1$
variables where the extra variable is given by the negative of the sum of $n$ variables.\footnote{See also Equation 6.4 in \cite{Murota:SIAM:2003} and Equation 29 in \cite{murota2021survey}.}
Hence,  we conclude that $\tilde{\Xi}$ is M-convex.

Therefore, given the M-convex policy goal $\tilde{\Xi}$, TTC-M in the hypothetical market satisfies the policy goal $\tilde \Xi$ (i.e., the policy goal given by $\Xi \equiv \Xi(f,f(\mu_0))$), constrained efficiency, individual rationality, and strategy-proofness.

We note that the outcome of our TTC algorithm is isomorphic to the outcome of TTC-M in the hypothetical market in the following sense. Student $s$ with type $t$ is allocated to $c$ for some $c \in \calc$ under preference profile $P=(P_s)_{s \in \S}$ at the outcome of the TTC algorithm if and only if student $s$ is allocated to the school-type pair $(c,t) \in \calc \times \calt$ under preference profile $\tilde P=(\tilde P_s)_{s \in \S}$ at TTC-M in the hypothetical market. Moreover, student $s$ is unassigned under preference profile $P=(P_s)_{s \in \S}$ at the outcome of the TTC algorithm if and only if student $s$ is allocated to $(\emptyset,\emptyset)$ under preference profile $\tilde P=(\tilde P_s)_{s \in \S}$ at TTC-M in the hypothetical market. Moreover, it is not difficult to confirm that our TTC's permissibility condition given in Section \ref{ttc_sec} is equivalent to TTC-M's acceptability condition.\footnote{To see this, consider a student $s\in \cals$ with type $t\in \calt$ at a matching $\mu$, (i) if $\mu_0(s)=c\in \calc$, then $s$ is permissible to $(c',t') \in \left(\calc \times \calt\right) \cup \{(\emptyset,\emptyset)\}$ if $f(\xi(\mu)+\chi_{c'}^{t'}-\chi_c^t) \geq f(\xi(\mu_0))$, which is the identical condition for acceptability, and (ii) if $\mu_0(s)=\emptyset$, then $s$ is permissible to $(c',t')\in \left(\calc \times \calt\right) \cup \{(\emptyset,\emptyset)\}$ if $f(\xi(\mu)+\chi_{c'}^{t'}) \geq f(\xi(\mu_0))$, which is equivalent to acceptability since $\chi_{\emptyset}^t$ is absent in our formulation.} The rest of the proof is devoted to showing that our TTC mechanism satisfies the desired properties in the original matching problem.

The result that the TTC mechanism weakly improves the distributional objective follows from the result
that the distribution corresponding to the TTC-M outcome is in  $\tilde{\Xi}$. Suppose that $\mu$ is the outcome of the TTC mechanism. Then, $(\xi(\mu),|\mathcal S|- n(\xi(\mu)) \in  \tilde{\Xi}$
 implies that $f(\mu) \ge f(\mu_0)$. Therefore, the TTC mechanism weakly improves the
distributional objective $f$.

To show constrained efficiency, let $\mu$ be the outcome of the TTC mechanism. Suppose, for contradiction, that there exists a matching $\mu'$ with $(\xi(\mu'),|\mathcal S|- n(\xi(\mu'))) \in \tilde{\Xi}$ that Pareto dominates matching $\mu$: for each student $s \in \S$, $\mu'(s) \mathrel{R_s} \mu(s)$, with at least one relation being strict. Then, by the construction of preferences $\mathrel{\tilde R_s}$ in the hypothetical market, we have $(\mu'(s), \tau'_s) \mathrel{\tilde R_s} (\mu(s),\tau''_s)$ for every student $s \in \S$, where (i) $\tau'_s=\tau_s$ if $\mu'(s) \in \calc$ and $\tau'_s=\emptyset$ otherwise, and (ii) $\tau''_s=\tau_s$ if $\mu(s) \in \calc$ and $\tau''_s=\emptyset$ otherwise. Moreover, one relation needs to be strict. However, this implies that, in the hypothetical market, $\{(s,(\mu'(s),\tau_s))\mid\mu'(s) \in \calc\} \cup \{(s, (\emptyset,\emptyset)) \mid \mu'(s) = \emptyset \}$ Pareto dominates $\{(s,(\mu(s),\tau_s))\mid\mu(s) \in \calc\} \cup \{(s, (\emptyset,\emptyset)) \mid \mu(s) = \emptyset \}$, while both matchings inducing distributions in the policy goal $\tilde \Xi$. This is a contradiction to the result that TTC-M is constrained efficient.

To show individual rationality, let matching $\mu$ be the outcome of the TTC mechanism and hence let $\{(s,(\mu(s),\tau_s))\mid\mu(s) \in \calc\} \cup \{(s, (\emptyset,\emptyset)) \mid \mu(s) = \emptyset \}$ be the result of TTC-M in the hypothetical market.  TTC-M is individually rational, hence, for each $s$ with $\mu_0(s) \in \calc$, we have $(\mu(s), \tau'_s) \mathrel{\tilde R_s} (\mu_0(s), \tau_s)$, where $\tau'_s=\tau_s$ if $\mu(s) \in \calc$ and $\tau'_s=\emptyset$, otherwise. By the construction of $\tilde R_s$, this relation implies  $\mu(s) \mathrel{R_s} \mu_0(s)$ for each $s$ with $\mu_0(s) \in \calc$. Moreover, for each $s$ with $\mu_0(s) = \emptyset$, we have $(\mu(s), \tau'_s) \mathrel{\tilde R_s} (\emptyset,\emptyset)$ where $\tau'_s=\tau_s$ if $\mu(s) \in \calc$ and $\tau'_s=\emptyset$, otherwise. These together imply that $\mu$ is individually rational in the original matching problem.

To show strategy-proofness, in the original matching problem, let $s$ be a student, $t$ her type, $P_{-s}$ the preference profile of students other than student $s$,
$P_s$ the true preference of student $s$, and $P'_s$ another preference of student $s.$ Furthermore, let $\mu(s)$ and $\mu'(s)$ be the assignments of student $s$
under $(P_s,P_{-s})$ and $(P_s',P_{-s})$ for TTC, respectively. Note that the previous argument establishes that, in the hypothetical market,
student $s$ is allocated to $(\mu(s),\tau'_s)$ and $(\mu'(s),\tau''_s)$, (where (i) $\tau'_s=t$ if $\mu(s) \in \calc$ and $\tau'_s=\emptyset$, otherwise, and
(ii) $\tau''_s=t$ if $\mu'(s) \in \calc$ and $\tau''_s=\emptyset$ otherwise) under $(\tilde P_s, \tilde P_{-s})$ and $(\tilde P'_s, \tilde P_{-s})$, respectively.
Because TTC-M is strategy-proof, it follows that $(\mu(s),\tau_s) \mathrel{\tilde R_s} (\mu'(s),\tau''_s)$. By the construction of $\tilde R_s$, this relation implies $\mu(s) \mathrel{R_s} \mu'(s)$, establishing strategy-proofness of the TTC mechanism.
\end{proof}
\medskip

\begin{proof}[Proof of Theorem \ref{thm:characterization}]

We first prove that $\Xi(f,\lambda)$ is M$^{\natural}$-convex for every $\lambda\in \mathbb{R}$ if and only if $f$ is pseudo M$^{\natural}$-concave.
\medskip

\noindent
\emph{The if direction:} Suppose that $f$ is pseudo M$^{\natural}$-concave and fix $\lambda\in \mathbb{R}$. Consider $\xi, \tilde{\xi} \in \Xi(f,\lambda)$ and $(c,t)\in \calc \times \calt$ with $\xi_c^t>\tilde{\xi}_c^t \,$.
By definition, $f(\xi),f(\tilde{\xi}) \geq \lambda$.
By pseudo M$^{\natural}$-concavity, we have, either (i) $\min \{f(\xi-\chi_c^t), f(\tilde{\xi}+\chi_c^t)\} \geq \min \{f(\xi),f(\tilde{\xi})\}$, or
(ii) there exists $(c',t')\in \calc \times \calt$ with $\xi_{c'}^{t'}<\tilde{\xi}_{c'}^{t'}$ such that
  \begin{center}
  $\min \{f(\xi-\chi_c^t+\chi_{c'}^{t'}), f(\tilde{\xi}+\chi_c^t-\chi_{c'}^{t'})\} \geq \min \{f(\xi),f(\tilde{\xi})\}$.
  \end{center}

Let us consider case (i) first. In this case, since $f(\xi),f(\tilde{\xi}) \geq \lambda$, we have $f(\xi-\chi_c^t), f(\tilde{\xi}+\chi_c^t)  \geq \lambda$. Hence, we get
$\xi-\chi_c^t, \tilde{\xi}+\chi_c^t \in \Xi(f,\lambda)$.

Next, let us consider case (ii). In this case, since $f(\xi),f(\tilde{\xi}) \geq \lambda$, we have $f(\xi-\chi_c^t+\chi_{c'}^{t'}),f(\tilde{\xi}+\chi_c^t-\chi_{c'}^{t'}) \geq \lambda$. Hence, we have $\xi-\chi_c^t+\chi_{c'}^{t'},\tilde{\xi}+\chi_c^t-\chi_{c'}^{t'}\in \Xi(f,\lambda)$.

We conclude that either $\xi-\chi_c^t, \tilde{\xi}+\chi_c^t \in \Xi(f,\lambda)$ or there exists $(c',t')\in \calc \times \calt$ with $\xi_{c'}^{t'}<\tilde{\xi}_{c'}^{t'}$
such that $\xi-\chi_c^t+\chi_{c'}^{t'},\tilde{\xi}+\chi_c^t-\chi_{c'}^{t'}\in \Xi(f,\lambda)$, implying that $\Xi(f,\lambda)$ is M$^{\natural}$-convex.

\medskip
\noindent
\emph{The only-if direction:} Suppose that $\Xi(f,\lambda)$ is M$^{\natural}$-convex for every $\lambda\in \mathbb{R}$.
Assume, for contradiction, that $f$ is not pseudo M$^{\natural}$-concave. Therefore, there exist
$\xi,\tilde{\xi} \in \Xi^0$ and $(c,t)\in \calc \times \calt$ with $\xi_c^t>\tilde{\xi}_c^t \,$ such that
(i) either $\xi-\chi_c^t \notin \Xi^0$ or $\tilde{\xi}+\chi_c^t \notin \Xi^0$ or $\min \{f(\xi-\chi_c^t), f(\tilde{\xi}+\chi_c^t)\} < \min \{f(\xi),f(\tilde{\xi})\}$
and (ii) for every $(c',t')\in \calc \times \calt$ with $\xi_{c'}^{t'}<\tilde{\xi}_{c'}^{t'}$ we have either
$\xi-\chi_c^t+\chi_{c'}^{t'} \notin \Xi^0$ or $\tilde{\xi}+\chi_c^t-\chi_{c'}^{t'} \notin \Xi^0$ or
$\min \{f(\xi-\chi_c^t+\chi_{c'}^{t'}), f(\tilde{\xi}+\chi_c^t-\chi_{c'}^{t'})\} < \min \{f(\xi),f(\tilde{\xi})\}$. Now, let $\lambda \equiv \min\{f(\xi),f(\tilde{\xi})\}$. These conditions imply that $\Xi(f,\lambda)$ is not M$^{\natural}$-convex because for M$^{\natural}$-convexity we need either (1) $\xi-\chi_c^t, \tilde{\xi}+\chi_c^t \in \Xi(f,\lambda)$ or (2) there exists $(c',t')\in \calc \times \calt$ with $\xi_{c'}^{t'}<\tilde{\xi}_{c'}^{t'}$ such that
$\xi-\chi_c^t+\chi_{c'}^{t'},\tilde{\xi}+\chi_c^t-\chi_{c'}^{t'}\in \Xi(f,\lambda)$. The first condition cannot hold because of (i) and the second condition
cannot hold because of (ii). Therefore, we get a contradiction to M$^{\natural}$-convexity of $\Xi(f,\lambda)$.
\medskip

Next, we prove the following: If $\Xi\subseteq \Xi^0$ is M$^{\natural}$-convex, then there exist a pseudo M$^{\natural}$-concave function
$f$ and a constant $\lambda \in \mathbb{R}$ such that $\Xi(f,\lambda)=\Xi$.

Let $f=f^{\Xi}_D$, where the distributional objective $f^{\Xi}_D$ is defined using the discrete metric in Section \ref{sec:alternative}:
$f^{\Xi}_D(\xi)=1$ if $\xi \in \Xi$ and $f^{\Xi}_D(\xi)=0$ otherwise (i.e., if $\xi\in \Xi^0 \setminus \Xi$). Since
$\Xi \subseteq \Xi^0$ is M$^{\natural}$-convex, Lemma \ref{lem:policyobjective} implies that
$f^{\Xi}_D$ is pseudo M$^{\natural}$-concave.

Finally, we show that $\Xi(f^{\Xi}_D,\lambda) = \Xi  $ for $\lambda=1$. For every $\xi \in \Xi(f^{\Xi}_D,1)$,
$f^{\Xi}_D(\xi)=1$, which implies that $\xi \in \Xi $ by the construction of $f^{\Xi}_D$. Therefore,
$\Xi(f^{\Xi}_D,1)  \subseteq \Xi$. Now, let $\xi \in \Xi $. Then, by the construction of $f^{\Xi}_D$,
$f^{\Xi}_D(\xi)=1$, so $\xi \in \Xi(f^{\Xi}_D,1) $. Therefore, $\Xi \subseteq \Xi(f^{\Xi}_D,1)$.
We conclude that $\Xi(f^{\Xi}_D,1)=\Xi$.
\end{proof}
\medskip

\begin{proof}[Proof of Lemma \ref{lem:cheb}]
\emph{The if direction:} Suppose that $f_C^{\Xi}$ is pseudo M$^{\natural}$-concave. We need to show that $\Xi$ is M$^{\natural}$-convex. By Theorem \ref{thm:characterization},
for every $\lambda \in \mathbb{R}$, $\Xi(f_C^{\Xi},\lambda)$ is M$^{\natural}$-convex. Let $\lambda=0$. By definition of the Chebyshev distance,
$\Xi(f_C^{\Xi},0)=\Xi$. Therefore, $\Xi$ is M$^{\natural}$-convex.
\medskip

\noindent
\emph{The only-if direction:} Suppose that $\Xi$ is M$^{\natural}$-convex. We need to show that $f_C^{\Xi}$ is pseudo M$^{\natural}$-concave.
By Theorem \ref{thm:characterization}, it suffices to show that $\Xi(f_C^{\Xi},\lambda)$ is M$^{\natural}$-convex for all $\lambda \leq 0$.\footnote{The reason is that, by definition, $f_C^{\Xi}$ only takes non-positive values.}

Since the Chebychev distance $D_C$ only takes integer values, we have $\Xi(f_C^{\Xi},\lambda') = \Xi(f_C^{\Xi},\lambda)$ for each $\lambda \in \mathbb{Z}_-$ and
$\lambda' \in  \mathbb{R}$ with $\lambda - \lambda' \in (0,1)$.\footnote{$\mathbb{Z_-}$ denotes the set of non-positive integers including zero}.
Hence, it suffices to show that
$\Xi(f_C^{\Xi},\lambda)$ is M$^{\natural}$-convex for each $\lambda \in \mathbb{Z_-}$. We prove this claim by mathematical induction. For the base case, when $\lambda=0$, we have $\Xi(f_C^{\Xi},0) = \Xi$, which is M$^{\natural}$-convex by assumption. In what follows, we prove the inductive step
that if $\Xi(f_C^{\Xi},\lambda)$ is M$^{\natural}$-convex for a $\lambda \in \mathbb{Z}_-$, then $\Xi(f_C^{\Xi},\lambda-1)$ is also M$^{\natural}$-convex.

Consider $\xi, \tilde{\xi} \in \Xi(f_C^{\Xi},\lambda-1)$ and $(c,t) \in \calc \times \calt$ with $\xi_c^t>\tilde{\xi}_c^t$ . First, note that $\xi \in \Xi(f_C^{\Xi},\lambda-1)$ implies that there exists $\dot{\xi} \in \Xi(f_C^{\Xi},\lambda)$ such that $D_C(\xi,\dot{\xi}) \leq 1$. Similarly, $\tilde{\xi} \in \Xi(f_C^{\Xi},\lambda-1)$ implies that there exists $\ddot{\xi} \in \Xi(f_C^{\Xi},\lambda)$ such that $D_C(\tilde{\xi},\ddot{\xi}) \leq 1$.

We consider the following two cases that are exhaustive.

Case 1 is given by the following condition: ``there exists $\dot{\xi} \in \Xi(f_C^{\Xi},\lambda)$ with $D_C(\xi,\dot{\xi}) \leq 1$ and $\xi_c^t - \dot{\xi}_c^t \in \{0,1\}$'' and ``there exists $\ddot{\xi} \in \Xi(f_C^{\Xi},\lambda)$ with $D_C(\tilde{\xi},\ddot{\xi}) \leq 1$ and $\tilde{\xi}_c^t - \ddot{\xi}_c^t \in \{0,-1\}$.''

Case 2 is given by the following condition: ``for all $\dot{\xi} \in \Xi(f_C^{\Xi},\lambda)$ with $D_C(\xi,\dot{\xi}) \leq 1$, we have $\xi_c^t - \dot{\xi}_c^t = -1$'' or ``for all $\ddot{\xi} \in \Xi(f_C^{\Xi},\lambda)$ with $D_C(\tilde{\xi},\ddot{\xi}) \leq 1$, we have $\tilde{\xi}_c^t - \ddot{\xi}_c^t = 1$.''

\medskip
\textbf{Case 1:} Take arbitrary $\dot{\xi}$ and $\ddot{\xi}$ that satisfy the conditions described in Case 1. Since we have $\xi_c^t - \dot{\xi}_c^t \in \{0, 1\}$, we can conclude that $D_C(\xi-\chi_c^t,\dot{\xi})=\max\{|(\xi-\chi_c^t)^t_c-(\dot{\xi})^t_c|, (|(\xi-\chi_c^t)^{t'}_{c'}-(\dot{\xi})^{t'}_{c'})|)_{(c',t') \neq (c,t)}\} \le 1$.
Similarly, since we have $\tilde{\xi}_c^t - \ddot{\xi}_c^t \in \{-1, 0\}$, we can conclude that $D_C(\tilde{\xi}+\chi_c^t,\ddot{\xi}) = \max\{|(\tilde{\xi}+\chi_c^t)^t_c-(\ddot{\xi})^t_c|, (|(\tilde{\xi}+\chi_c^t)^{t'}_{c'}-(\ddot{\xi})^{t'}_{c'})|)_{(c',t') \neq (c,t)}\} \le 1$. Hence, $\xi-\chi_c^t \in \Xi(f_C^{\Xi},\lambda-1)$ and $\tilde{\xi}+\chi_c^t \in \Xi(f_C^{\Xi},\lambda-1)$.

We consider Case 2 next.

\medskip
\textbf{Case 2:} Take arbitrary $\dot{\xi}$ and $\ddot{\xi}$ that satisfy the conditions described in Case 2. Note that, since $\xi_c^t>\tilde{\xi}_c^t$, and either $\dot{\xi}_c^t > \xi_c^t$ or  $\tilde{\xi}_c^t > \ddot{\xi}_c^t$, we have $\dot{\xi}_c^t>\ddot{\xi}_c^t$. Also note that, the distances between the respective distributions are bounded above by $1$.

Since $\Xi(f_C^{\Xi},\lambda)$ is M$^{\natural}$-convex, $\dot{\xi}, \ddot{\xi} \in \Xi(f_C^{\Xi},\lambda)$ and $\dot{\xi}_c^t>\ddot{\xi}_c^t$, we have

\begin{enumerate}[(i)]
\item $\dot{\xi}-\chi_c^t \in \Xi(f_C^{\Xi},\lambda)$ and $\ddot{\xi}+\chi_c^t \in \Xi(f_C^{\Xi},\lambda)$, or
\item there exist school $c'$ and type $t'$  with $\dot{\xi}_{c'}^{t'}<\ddot{\xi}_{c'}^{t'}$ such that
$\dot{\xi}-\chi_c^t+\chi_{c'}^{t'} \in \Xi(f_C^{\Xi},\lambda)$ and $\ddot{\xi}+\chi_c^t-\chi_{c'}^{t'}\in \Xi(f_C^{\Xi},\lambda)$.
\end{enumerate}

If (i) holds, then since $D_C(\dot{\xi}-\chi_c^t,\xi-\chi_c^t)=D_C(\dot{\xi},\xi) \leq 1$
and $D_C(\ddot{\xi}+\chi_c^t,\tilde{\xi}+\chi_c^t)=D_C(\ddot{\xi},\tilde{\xi}) \leq 1$, we have $\xi-\chi_c^t \in \Xi(f_C^{\Xi},\lambda-1)$ and $\tilde{\xi}+\chi_c^t \in \Xi(f_C^{\Xi},\lambda-1)$.

If (ii) holds, then we analyze two contingencies: (a) $\xi_{c'}^{t'}<\tilde{\xi}_{c'}^{t'}$ and (b) $\xi_{c'}^{t'} \geq \tilde{\xi}_{c'}^{t'}$.

For contingency (a), since $D_C(\dot{\xi}-\chi_c^t+\chi_{c'}^{t'} ,\xi-\chi_c^t+\chi_{c'}^{t'}) \leq 1$ and $D_C(\ddot{\xi}+\chi_c^t-\chi_{c'}^{t'},\tilde{\xi}+\chi_c^t-\chi_{c'}^{t'}) \leq 1$, we have $\xi-\chi_c^t +\chi_{c'}^{t'}\in \Xi(f_C^{\Xi},\lambda-1)$ and $\tilde{\xi}+\chi_c^t-\chi_{c'}^{t'} \in \Xi(f_C^{\Xi},\lambda-1)$ where $\xi_{c'}^{t'}<\tilde{\xi}_{c'}^{t'}$.

For contingency (b), for $\dot{\xi}_{c'}^{t'}<\ddot{\xi}_{c'}^{t'}$ and $\xi_{c'}^{t'} \geq \tilde{\xi}_{c'}^{t'}$ to hold simultaneously, we need to have one of the following three subcases: (1) $\dot{\xi}_{c'}^{t'} = \xi_{c'}^{t'}-1$ and $\ddot{\xi}_{c'}^{t'} = \tilde{\xi}_{c'}^{t'}+1$, or
(2) $\dot{\xi}_{c'}^{t'} = \xi_{c'}^{t'}$ and $\ddot{\xi}_{c'}^{t'} = \tilde{\xi}_{c'}^{t'}+1$ or (3) $\dot{\xi}_{c'}^{t'} = \xi_{c'}^{t'}-1$ and $\ddot{\xi}_{c'}^{t'} = \tilde{\xi}_{c'}^{t'}$.

In the first subcase b(1), we have $D_C(\dot{\xi}-\chi_c^t+\chi_{c'}^{t'}, \xi-\chi_c^t) = D_C(\dot{\xi}+\chi_{c'}^{t'}, \xi)$. Moreover, since $\dot{\xi}_{c'}^{t'} = \xi_{c'}^{t'}-1$, we get $D_C(\dot{\xi}+\chi_{c'}^{t'}, \xi) \le D_C(\dot{\xi}, \xi) \le 1$. Therefore, since $\dot{\xi}-\chi_c^t+\chi_{c'}^{t'} \in \Xi(f_C^{\Xi},\lambda)$, we conclude $\xi-\chi_c^t \in \Xi(f_C^{\Xi},\lambda-1)$.
Similarly, we have $D_C(\ddot{\xi}+\chi_c^t-\chi_{c'}^{t'}, \tilde{\xi}+\chi_c^t) =D_C(\ddot{\xi}-\chi_{c'}^{t'}, \tilde{\xi})$. Moreover, since $\ddot{\xi}_{c'}^{t'} = \tilde{\xi}_{c'}^{t'}+1$, we get $D_C(\ddot{\xi}-\chi_{c'}^{t'}, \tilde{\xi}) \le D_C(\ddot{\xi},\tilde{\xi}) \le 1$. Therefore, since $\ddot{\xi}+\chi_c^t-\chi_{c'}^{t'}\in \Xi(f_C^{\Xi},\lambda)$, we conclude $\tilde{\xi}+\chi_c^t \in \Xi(f_C^{\Xi},\lambda-1)$.


In the second subcase b(2), $D_C(\ddot{\xi}+\chi_c^t-\chi_{c'}^{t'},\tilde{\xi}+\chi_c^t) \leq D_C(\ddot{\xi},\tilde{\xi}) = 1$. Hence, $\tilde{\xi}+\chi_c^t \in \Xi(f_C^{\Xi},\lambda-1)$. Moreover, $\xi-\chi_c^t \in \Xi(f_C^{\Xi},\lambda-1)$ unless $\dot{\xi}_c^t=\xi_c^t+1$. When $\dot{\xi}_c^t=\xi_c^t+1$ and $\dot{\xi}_{c'}^{t'} = \xi_{c'}^{t'}$, however, we have $D_C(\dot{\xi}-\chi_c^t+\chi_{c'}^{t'}, \xi-\chi_c^t) = 1$\, so $\xi-\chi_c^t \in \Xi(f_C^{\Xi},\lambda-1)$. This is because, the absolute value of the differences between these
two distributions is one at $(t,c)$ and one at $(t',c')$ (for all other $(t'',c'')$, it is equal to the corresponding difference between $\xi$ and $\dot{\xi}$, which is at most one).

In the third subcase b(3), $D_C(\dot{\xi}-\chi_c^t+\chi_{c'}^{t'}, \xi-\chi_c^t) \leq D_C(\dot{\xi}, \xi) = 1$. Hence, $\xi-\chi_c^t \in \Xi(f_C^{\Xi},\lambda-1)$.
Moreover, $\tilde{\xi}+\chi_c^t \in \Xi(f_C^{\Xi},\lambda-1)$ unless $\ddot{\xi}_c^t=\tilde \xi_c^t-1$. When $\ddot{\xi}_c^t=\tilde \xi_c^t-1$ and $\ddot{\xi}_{c'}^{t'} = \tilde{\xi}_{c'}^{t'}$, however, we have $D_C(\ddot{\xi}+\chi_c^t-\chi_{c'}^{t'},\tilde{\xi}+\chi_c^t) = 1$. So, $\tilde \xi - \chi_c^t \in \Xi(f_C^{\Xi},\lambda-1)$. This is because, the absolute value of the differences between these two distributions is one at
$(t,c)$ and one at $(t',c')$ (for all other $(t'',c'')$, it is equal to the corresponding difference between $\tilde{\xi}$ and $\ddot{\xi}$, which is at most one).

This completes the proof that $\Xi(f_C^{\Xi},\lambda-1)$ is M$^{\natural}$-convex.

Hence, by mathematical induction, we have proven that  $\Xi(f_C^{\Xi},\lambda)$ is M$^{\natural}$-convex for each $\lambda \in \mathbb{Z}_-$, and therefore for each $\lambda \leq 0$. Therefore, $f_C^{\Xi}$ is pseudo M$^{\natural}$-concave by Theorem \ref{thm:characterization}.
\end{proof}

\medskip

\begin{proof}[Proof of Lemma \ref{lem:policyobjective}]
\emph{The if direction:} Suppose that $f^{\Xi}_D$ is pseudo M$^{\natural}$-concave. By Theorem \ref{thm:characterization}, for every $\lambda\in \mathbb{R}$,
$\Xi(f^{\Xi}_D,\lambda)$ is M$^{\natural}$-convex. Let $\lambda=1$, then $\Xi(f^{\Xi}_D,\lambda)=\Xi$. Therefore, $\Xi$ is M$^{\natural}$-convex.

\medskip
\noindent
\emph{The only-if direction:} Suppose that the policy goal $\Xi \subseteq \Xi^0$ is M$^{\natural}$-convex. To show that $f^{\Xi}_D$ is pseudo M$^{\natural}$-concave,
we show, for every $\lambda \in \mathbb{R}$, $\Xi(f^{\Xi}_D,\lambda)$ is M$^{\natural}$-convex.

If $\lambda>1$, then $\Xi(f^{\Xi}_D,\lambda)=\emptyset$, which is trivially M$^{\natural}$-convex. If
$1\geq \lambda>0$, then $\Xi(f^{\Xi}_D,\lambda)=\Xi$, which is M$^{\natural}$-convex by assumption.
Finally, if $\lambda \leq 0$, then $\Xi(f^{\Xi}_D,\lambda)=\Xi^0$, which we show to be M$^{\natural}$-convex.

Suppose that there exist $\xi, \tilde{\xi}\in \Xi^0$ and $(c,t) \in \calc \times \calt$ with $\xi_c^t>\tilde{\xi}_{c}^{t}$.
To show M$^{\natural}$-convexity, we analyze two possible cases depending on whether
$n_c(\tilde \xi) < q_c$ or $n_c(\tilde \xi) = q_c$.

\textbf{Case 1:} If $n_c(\tilde \xi) < q_c$, then both $\xi-\chi_c^t$ and $\tilde{\xi}+\chi_c^t$ satisfy the capacity constraint
at school $c$. Therefore, $\xi-\chi_c^t \in \Xi^0$ and $\tilde{\xi}+\chi_c^t \in \Xi^0$.

\textbf{Case 2:} If $n_c(\tilde \xi) = q_c$, then there exists $t'\in \calt$ such that $\xi_c^{t'} < \tilde{\xi}_{c}^{t'}$. This is because, otherwise, $\xi$ cannot satisfy the capacity constraint at school $c$ given that
$\xi_c^t>\tilde{\xi}_{c}^{t}$ and $n_c(\tilde \xi) = q_c$.
For every school, the number of students that it has in $\xi-\chi_c^t+\chi_{c}^{t'}$ and $\xi$ are the same, which implies that
$\xi-\chi_c^t+\chi_{c}^{t'}\in \Xi^0$ since $\xi \in \Xi^0$. Likewise, for every school, the number of students that it
has in $\tilde{\xi}+\chi_c^t-\chi_{c}^{t'}$ and $\tilde{\xi}$ are the same, so $\tilde{\xi}+\chi_c^t-\chi_{c}^{t'}\in \Xi^0$
since $\tilde{\xi}\in \Xi^0$.

Therefore, $\Xi^0$ is M$^{\natural}$-convex. Since all upper contour sets of $f^{\Xi}_D$ are M$^{\natural}$-convex,
we conclude that $f^{\Xi}_D$ is pseudo M$^{\natural}$-concave by Theorem \ref{thm:characterization}.
\end{proof}

\medskip

\begin{proof}[Proof of Proposition \ref{prop:quotas}]
We show that the school-level quota policy
\[\Xi^Q = \left \{\xi \in \Xi^0 \mathlarger{\mathlarger{\mid}}  u_c \geq \sum_{t} \xi_c^t \geq l_c \text{ for all } c\in \calc \right \}\]
is M$^{\natural}$-convex.
For a distribution $\xi \in \mathbb N^{|\C|\times |\T|}$, let us call
\begin{enumerate}
\item $q_c \geq \sum_t \xi_c^t$ (equivalently, $q_c \geq n_c(\xi)$) the \emph{capacity constraint} for school $c\in \calc$,
\item $u_c \geq \sum_{t} \xi_c^t$ (equivalently, $u_c \geq n_c(\xi)$) the \emph{upper quota constraint} for school $c \in \calc$, and
\item $\sum_{t} \xi_c^t \geq l_c$ (equivalently, $n_c(\xi) \geq l_c$) the \emph{lower quota constraint} for school $c\in \calc$.
\end{enumerate}

Suppose that there exist $\xi,\tilde{\xi}\in \Xi^Q$ and $(c,t)\in \calc \times \calt$ with $\xi_c^t>\tilde{\xi}_{c}^{t}$. To show M$^{\natural}$-convexity,
we analyze two possible cases depending on whether $\tilde{\xi}_{c}^{t'} \le \xi_{c}^{t'}$ for all $t' \in \calt$ (Case 1) or there exists
$t' \in \calt$ such that $\tilde{\xi}_{c}^{t'} > \xi_{c}^{t'}$ (Case 2).

\medskip
\textbf{Case 1:}  Suppose $\tilde{\xi}_{c}^{t'} \le \xi_{c}^{t'}$ for all $t' \in \calt$. In this case, we first show
$\xi-\chi_c^t \in \Xi^Q$ and, then, $\tilde{\xi}+\chi_c^t \in \Xi^Q$.

\smallskip
\noindent \emph{Capacity constraints for $\xi-\chi_c^t$:} For each school $c'\in \calc$, $n_{c'}(\xi-\chi_c^t) \leq n_{c'}(\xi)$. Therefore,
since $\xi$ satisfies the quota constraints at all schools, so does $\xi-\chi_c^t$.

\smallskip
\noindent \emph{Upper quota constraints for $\xi-\chi_c^t$:} Since $n_{c'}(\xi-\chi_c^t) \leq n_{c'}(\xi)$ for each school $c'\in \calc$ and
$\xi$ satisfies the upper quota constraints at all schools, $\xi-\chi_c^t$ also satisfies the upper quota constraints at all schools.

\smallskip
\noindent \emph{Lower quota constraints for $\xi-\chi_c^t$:} Notice that $\xi_{c}^{t'} \geq \tilde{\xi}_{c}^{t'}$ for all $t' \in \calt$ and
$\xi_c^t>\tilde{\xi}_{c}^{t}$ imply $n_c(\tilde \xi) < n_c(\xi)$. As a consequence, together with $n_c(\xi-\chi_c^t)=n_c(\xi)-1$,
we get $n_c(\xi-\chi_c^t)\geq n_c(\tilde \xi)$. Since $\tilde \xi$ satisfies the lower quota constraint for school $c$, so does $\xi-\chi_c^t$.
For $c'\in \calc \setminus \{c\}$, $n_{c'}(\xi-\chi_c^t) = n_{c'}(\xi)$. Therefore, $\xi-\chi_c^t$ satisfies the lower quota constraint at
$c'$ because $\xi$ satisfies the same constraint.

\smallskip
\noindent \emph{Capacity constraints for $\tilde \xi+\chi_c^t$:} For each school $c' \in \calc \setminus \{c\}$, $n_{c'}(\tilde \xi+\chi_c^t)=n_{c'}(\tilde \xi)$.
Since $\tilde \xi$ satisfies the quota constraint at school $c'$, so does $\tilde \xi-\chi_c^t$. Furthermore, $\xi_{c}^{t'} \geq \tilde{\xi}_{c}^{t'}$ for all
$t' \in \calt$ and $\xi_c^t>\tilde{\xi}_{c}^{t}$ yield $n_c(\tilde \xi) < n_c(\xi)$. This inequality, together with $n_{c}(\tilde \xi+\chi_c^t)=n_{c}(\tilde \xi)+1$,
yields $n_{c}(\tilde \xi+\chi_c^t) \leq n_c(\xi)$. Therefore, since $\xi$ satisfies the capacity constraint at school $c$, so does
$\tilde \xi + \chi_c^t$.

\smallskip
\noindent \emph{Upper quota constraints for $\tilde \xi+\chi_c^t$:} We show in the previous paragraph that $n_{c}(\tilde \xi+\chi_c^t) \leq n_c(\xi)$
and, for each school $c' \in \calc \setminus \{c\}$, $n_{c'}(\tilde \xi+\chi_c^t)=n_{c'}(\tilde \xi)$. Therefore, $\tilde \xi+\chi_c^t$
satisfies the upper quota constraints at all schools because $\xi$ and $\tilde \xi$ satisfy them.

\smallskip
\noindent \emph{Lower quota constraints for $\tilde \xi+\chi_c^t$:} For every school $c'$, we have $n_{c'}(\tilde \xi+\chi_c^t)=n_{c'}(\tilde \xi)+1$. Furthermore,
$\tilde \xi$ satisfies the lower quota constraints for all schools. Hence, $\tilde \xi+\chi_c^t$ satisfies the lower quota constraints for all schools as well.

\smallskip
Therefore, we conclude this case by concluding that $\xi-\chi_c^t \in \Xi^Q$ and $\tilde{\xi}+\chi_c^t \in \Xi^Q$.

\medskip
\textbf{Case 2:} Suppose that $t'\in \calt$ is such that $\tilde{\xi}_{c}^{t'} > \xi_{c}^{t'}$. In this case
we show that $\xi-\chi_c^t+\chi_{c}^{t'} \in \Xi^Q$ and $\tilde{\xi}+\chi_c^t-\chi_{c}^{t'}\in \Xi^Q$.

For every $\hat c \in \calc$, $n_{\hat c}(\xi-\chi_c^t+\chi_{c}^{t'}) = n_{\hat c}(\xi)$ and
$n_{\hat{c}}(\tilde{\xi}+\chi_c^t-\chi_{c}^{t'})=n_{\hat{c}}(\tilde \xi)$. Therefore, $\xi-\chi_c^t+\chi_{c}^{t'}$ satisfies
capacity constraints, lower quota constraints, and upper quota constraints at all schools because so does $\xi$. Likewise,
$\tilde{\xi}+\chi_c^t-\chi_{c}^{t'}$ satisfies capacity constraints, lower quota constraints, and upper quota constraints at all schools
because so does $\tilde \xi$. Therefore, $\xi-\chi_c^t+\chi_{c}^{t'} \in \Xi^Q$ and $\tilde{\xi}+\chi_c^t-\chi_{c}^{t'}\in \Xi^Q$.

We conclude that $\Xi^Q$ is an M$^{\natural}$-convex set. The desired conclusion then follows from M$^{\natural}$-convexity of $\Xi^Q$,
Theorem \ref{thm:ttc}, and Lemmas \ref{lem:cheb} and \ref{lem:policyobjective}.
\end{proof}
\medskip

\begin{proof}[Proof of Proposition \ref{prop:convexdiv}]
We show that the school-level diversity policy
\[\Xi^D = \left \{\xi \in \Xi^0 \mathlarger{\mathlarger{\mid}} q_c^t \geq \xi_c^t \geq p_c^t \text{ for all } (c,t)\in \calc \times \calt\right \}\]
is M$^{\natural}$-convex.

For a distribution $\xi \in \mathbb N^{|\C|\times |\T|}$, let us call
\begin{enumerate}
\item $q_c \geq \sum_t \xi_c^t$ (equivalently, $q_c \geq n_c(\xi)$) the \emph{capacity constraint} for school $c\in \calc$,
\item $q_c^t \geq \xi_c^t$ the \emph{ceiling constraint} for school $c\in \calc$ and type $t\in \calt$, and
\item $\xi_c^t \geq p_c^t$ the \emph{floor constraint} for school $c\in \calc$ and type $t\in \calt$.
\end{enumerate}

Suppose that there exist $\xi,\tilde{\xi}\in \Xi^D$ and $(c,t) \in \calc \times \calt$ with $\xi_c^t>\tilde{\xi}_{c}^{t}$.
To show M$^{\natural}$-convexity, we analyze two possible cases depending on whether $n_c(\tilde \xi) < q_c$ (Case 1)
or $n_c(\tilde \xi) = q_c$ (Case 2).

\medskip
\textbf{Case 1:} Suppose that $n_c(\tilde \xi) < q_c$. We show that $\xi-\chi_c^t \in \Xi^D$ and $\tilde{\xi}+\chi_c^t \in \Xi^D$.

\smallskip
\noindent \emph{Constraints for $\xi-\chi_c^t$:} Since $\xi$ satisfies the capacity constraints for all schools, so does $\xi-\chi_c^t$.
Likewise, since $\xi$ satisfies the ceiling constraints for all schools and types, so does $\xi-\chi_c^t$. If $(c',t')\in \calc \times \calt$ is
different than $(c,t)$, then $(\xi-\chi_c^t)_{c'}^{t'}=\xi_{c'}^{t'}$. Since $\xi$ satisfies the floor constraint for school $c'$ and
type $t'$, $\xi-\chi_c^t$ also satisfies it. For school $c$ and  type $t$, we have $(\xi-\chi_c^t)_{c}^{t}=\xi_c^t-1 \geq \tilde{\xi}_c^t$.
Therefore, $\xi-\chi_c^t$ satisfies the floor constraint for school $c$ and type $t$, because $\tilde \xi$ satisfies the constraint.

\smallskip
\noindent \emph{Constraints for $\tilde \xi + \chi_c^t$:} Since $n_c(\tilde \xi) < q_c$, $\tilde \xi + \chi_c^t$ satisfies the capacity constraint.
Furthermore, since $\tilde \xi$ satisfies the floor constraints for all schools and types,  $\tilde \xi + \chi_c^t$ also satisfies all
floor constraints. If $(c',t')\in \calc \times \calt$ is different than $(c,t)$, then $(\tilde \xi + \chi_c^t)_{c'}^{t'}=\tilde{\xi}_{c'}^{t'}$
implies that $\tilde \xi + \chi_c^t$ satisfies the ceiling constraint for school $c'$ and type $t'$ because $\tilde \xi$ satisfies the same
constraint. For school $c$ and type $t$, $(\tilde \xi + \chi_c^t)_{c}^{t}= \tilde{\xi}_c^t + 1 \leq \xi_c^t$. Therefore, $\tilde \xi + \chi_c^t$
satisfies the ceiling constraint for school $c$ and type $t$ because $\xi$ satisfies the constraint.

\medskip
\textbf{Case 2:} Suppose that $n_c(\tilde \xi) = q_c$. Then there exists $t'\in \calt$ such that $\xi_c^{t'} < \tilde{\xi}_{c}^{t'}$ because,
otherwise, $\xi$ cannot satisfy the capacity constraint (given that $\xi_c^t>\tilde{\xi}_{c}^{t}$ and $n(\tilde \xi) = q_c$). We show that
$\xi-\chi_c^t+\chi_{c}^{t'} \in \Xi^D$ and $\tilde{\xi}+\chi_c^t-\chi_{c}^{t'}\in \Xi^D$.

\smallskip
\noindent \emph{Constraints for $\xi-\chi_c^t+\chi_{c}^{t'}$:} For every school, the number of students assigned to the school in $\xi-\chi_c^t+\chi_{c}^{t'}$ and
the number of students assigned to the school in $\xi$ are the same. Therefore, $\xi-\chi_c^t+\chi_{c}^{t'}$ satisfies the capacity constraints for all schools
because $\xi$ also satisfies them. In addition, for any $(\hat c, \hat t) \in (\calc \times \calt) \setminus \{(c,t),(c,t')\}$, the
number of type-$\hat{t}$ students assigned to school $\hat c$ in $\xi-\chi_c^t+\chi_{c}^{t'}$ is the same as the
number of type-$\hat{t}$ students assigned to school $\hat c$ in $\xi$. Therefore, $\xi-\chi_c^t+\chi_{c}^{t'}$ satisfies the floor
and ceiling constraints for school $\hat c$ and type $\hat t$ because $\xi$ satisfies them. For school $c$ and type $t$, we have
\[\tilde{\xi}^t_c \leq \xi^t_c-1=(\xi-\chi_c^t+\chi_{c}^{t'})_c^t<\xi^t_c.\]
Therefore, $\xi-\chi_c^t+\chi_{c}^{t'}$ satisfies the floor constraint for school $c$ and type $t$ because $\tilde \xi$ satisfies the same constraint, and $\xi-\chi_c^t+\chi_{c}^{t'}$ satisfies the ceiling constraint for school $c$ and type $t$ because $\xi$ satisfies the same constraint. Similarly,  $\xi-\chi_c^t+\chi_{c}^{t'}$ satisfies the floor constraint for school $c$ and type $t'$ because $ \xi$ satisfies the same constraint, and $\xi-\chi_c^t+\chi_{c}^{t'}$ satisfies the ceiling constraint for school $c$ and type $t'$ because $\tilde \xi$ satisfies the same constraint.

\smallskip
\noindent \emph{Constraints for $\tilde \xi + \chi_c^t - \chi_{c}^{t'}$:} The proof is analogous as in the previous paragraph by changing the
roles of $t$ with $t'$ and $\xi$ with $\tilde \xi$.

\medskip

Hence, $\Xi^D$ is an M$^{\natural}$-convex set. The desired conclusion then follows from the fact that $\Xi^D$ is an M$^{\natural}$-convex set,
Theorem \ref{thm:ttc}, and Lemmas \ref{lem:cheb} and \ref{lem:policyobjective}.
\end{proof}
\medskip

\begin{proof}[Proof of Proposition \ref{prop:convexbal}]
We show that the exchange-feasiblity policy
\[\Xi^E = \left \{\xi \in \Xi^0 \mathlarger{\mathlarger{\mid}} \sum_{t,c:d(c)=d} \xi_c^t \geq k_{d} \text{ for all } d \in \cald \right \}\]
is M$^{\natural}$-convex.

For a distribution $\xi \in \mathbb N^{|\C|\times |\T|}$, let us call
\begin{enumerate}
\item $q_c \geq \sum_t \xi_c^t$ (equivalently, $q_c \geq n_c(\xi)$) the \emph{capacity constraint} for school $c\in \calc$,
\item $\sum_{t,c:d(c)=d} \xi_c^t \geq k_{d}$  the \emph{exchange-feasibility constraint} for district $d \in \cald.$
\end{enumerate}

Suppose that there exist $\xi,\tilde{\xi}\in \Xi^E$ and $(c,t) \in \calc \times \calt$ with $\xi_c^t>\tilde{\xi}_{c}^{t}$. Let us denote $d(c)$ by $d$.
To show M$^{\natural}$-convexity, we study three possible (non-disjoint, yet exhaustive) cases depending on the comparison between $n_d(\xi)$ and $k_{d}$,
and on the comparison between $n_c(\tilde \xi)$ and $q_c$: $n_d(\xi) > k_{d}$ and $n_c(\tilde \xi) < q_c$ (Case 1), $n_c(\tilde \xi) = q_c$ (Case 2), and $n_d(\xi) = k_{d}$ (Case 3).

\textbf{Case 1:} Suppose $n_d(\xi) > k_{d}$ and $n_c(\tilde \xi) < q_c$. We show that $\xi-\chi_c^t \in \Xi^E$ and $\tilde{\xi}+\chi_c^t \in \Xi^E$.

\smallskip
\noindent \emph{Constraints for $\xi-\chi_c^t$:} Since $\xi$ satisfies the capacity constraints for all schools, so does $\xi-\chi_c^t$.
Moreover, since $\xi$ satisfies the exchange-feasibility constraints for all districts and $n_d(\xi) > k_{d}$, so does $\xi-\chi_c^t$ (as $n_d(\xi-\chi_c^t) \geq k_{d}$).

\smallskip
\noindent \emph{Constraints for $\tilde \xi + \chi_c^t$:} Since $n_c(\tilde \xi) < q_c$, we have $n_c(\tilde \xi + \chi_c^t) \leq q_c$, and hence $\tilde \xi + \chi_c^t$ satisfies the capacity constraints for all schools. Furthermore, since $\tilde \xi$ satisfies the exchange-feasibility constraints for all districts,  $\tilde \xi + \chi_c^t$ clearly satisfies the exchange-feasibility constraints.

\textbf{Case 2:} Suppose $n_c(\tilde \xi) = q_c$, then we argue that there exist $t' \neq t$ such that $\xi_c^{t'} < \tilde{\xi}_{c}^{t'}$. This follows from the facts that $n_c(\tilde \xi)= q_c \geq n_c(\xi)$ and $\xi_c^t>\tilde{\xi}_{c}^{t}$. Then, both $\xi-\chi_c^t+\chi_{c}^{t'}$ and $\tilde{\xi}+\chi_c^t-\chi_{c}^{t'}$ clearly satisfy both the capacity constraints for all
schools and the exchange-feasibility constraints for all districts, as $n_c$ and $n_d$ values among $\xi$ and $\xi-\chi_c^t+\chi_{c}^{t'}$ remain the same, and $n_c$ and $n_d$ values among $\tilde \xi$ and $\tilde{\xi}+\chi_c^t-\chi_{c}^{t'}$ remain the same. Thus, $\xi-\chi_c^t+\chi_{c}^{t'} \in \Xi^E$ and $\tilde{\xi}+\chi_c^t-\chi_{c}^{t'}\in \Xi^E$.

\textbf{Case 3:} Suppose $n_d(\xi) = k_{d}$, for this case, we first note that $n_d(\xi) \leq n_d(\tilde \xi)$. Next, we consider two subcases of this case: $n_c(\xi) \le n_c(\tilde \xi)$ (Subcase 3.1), and $n_c(\xi) > n_c(\tilde \xi)$ (Subcase 3.2).

\textbf{Subcase 3.1:} Suppose $n_c(\xi) \le n_c(\tilde \xi)$, then similar to Case 2 above, it follows that there exists $t' \neq t$ such that $\xi_c^{t'} < \tilde{\xi}_{c}^{t'}$. Then one can verify that $\xi-\chi_c^t+\chi_{c}^{t'} \in \Xi^E$ and $\tilde{\xi}+\chi_c^t-\chi_{c}^{t'}\in \Xi^E$.

\textbf{Subcase 3.2:} Suppose $n_c(\xi) > n_c(\tilde \xi)$, then we first observe that there exists a $c' \neq c$ with $d(c')=d$ such that
$n_{c'}(\xi) < n_{c'}(\tilde \xi)$. This is because, otherwise, we cannot have $n_d(\xi) \leq n_d(\tilde \xi)$. Next, we observe that $n_{c'}(\xi) < n_{c'}(\tilde \xi)$
implies that $ \xi_{c'}^{t'} <  \tilde{\xi}_{c'}^{t'}$ for some $t'$. Finally, we argue that $\xi-\chi_c^t+\chi_{c'}^{t'} \in \Xi^E$ and $\tilde{\xi}+\chi_c^t-\chi_{c'}^{t'}\in \Xi^E$.

\smallskip
\noindent \emph{Exchange-feasibility constraints:} Both $\xi-\chi_c^t+\chi_{c'}^{t'}$ and $\tilde{\xi}+\chi_c^t-\chi_{c'}^{t'}$ clearly satisfy the exchange-feasibility constraints, $n_d$ values among $\xi$ and $\xi-\chi_c^t+\chi_{c}^{t'}$ remain the same, and $n_d$ values among $\tilde \xi$ and $\tilde{\xi}+\chi_c^t-\chi_{c}^{t'}$ remain the same.

\smallskip
\noindent \emph{Capacity constraints for $\xi-\chi_c^t+\chi_{c'}^{t'}$:} $\xi-\chi_c^t+\chi_{c'}^{t'}$ clearly satisfies the capacity constraint for $c$ and it also satisfies the capacity constraint for $c'$ since $n_{c'}(\xi) < n_{c'}(\tilde \xi) \leq q_{c'}$.  It also satisfies the capacity constraints for other schools since $\xi$ satisfies the capacity constraints for other schools.

\smallskip
\noindent \emph{Capacity constraints for $\tilde{\xi}+\chi_c^t-\chi_{c'}^{t'}$:} $\tilde{\xi}+\chi_c^t-\chi_{c'}^{t'}$ clearly satisfies the capacity constraint for $c'$
and it also satisfies the capacity constraint for $c$ since  $n_c(\tilde \xi) < n_c(\xi) \leq q_c$.  It also satisfies the capacity constraints for other schools since $\tilde \xi$ satisfies the capacity constraints for other schools.

Thus, $\xi-\chi_c^t+\chi_{c'}^{t'} \in \Xi^E$ and $\tilde{\xi}+\chi_c^t-\chi_{c'}^{t'}\in \Xi^E$.

Hence, we establish that $\Xi^E$ is M$^{\natural}$-convex.

The result then follows from Theorem \ref{thm:ttc}, and Lemmas \ref{lem:cheb} and \ref{lem:policyobjective} because $\Xi^E$ is M$^{\natural}$-convex.
\end{proof}
\medskip

\begin{proof}[Proof of Proposition \ref{prop:mix}]
We show that the combination of school-level diversity and exchange-feasibility policies
\[\Xi^{DE} = \left\{\xi \in \Xi^0 \mathlarger{\mathlarger{\mid}} q_c^t \geq \xi_c^t \geq p_c^t \text{ , } \forall (c,t)\in \calc \times \calt \mbox{ and }  \sum_{t,c:d(c) = d} \xi_c^t \geq k_{d} \text{ } \text{ , } \forall d \in \cald \right\}\]
is M$^{\natural}$-convex.

As in Propositions \ref{prop:convexdiv} and \ref{prop:convexbal}, for a distribution $\xi \in \mathbb N^{|\C|\times |\T|}$, let us call
\begin{enumerate}
\item $q_c \geq \sum_t \xi_c^t$ (equivalently, $q_c \geq n_c(\xi)$) the \emph{capacity constraint} for school $c\in \calc$,
\item $q_c^t \geq \xi_c^t$ the \emph{ceiling constraint} for school $c\in \calc$ and type $t\in \calt$
\item $\xi_c^t \geq p_c^t$ the \emph{floor constraint} for school $c\in \calc$ and type $t\in \calt$, and
\item $\sum_{t,c:d(c)=d} \xi_c^t \geq k_{d}$  the \emph{exchange-feasibility constraint} for district $d \in \cald.$
\end{enumerate}

Suppose that there exist $\xi,\tilde{\xi}\in \Xi^{DE}$ and $(c,t) \in \calc \times \calt$ with $\xi_c^t>\tilde{\xi}_{c}^{t}$. Let us denote $d(c)$ by $d$.
To show M$^{\natural}$-convexity, we investigate three possible (non-disjoint, yet exhaustive) cases depending on the comparison between $n_d(\xi)$ and $k_{d}$,
and on the comparison between $n_c(\tilde \xi)$ and $q_c$: $n_d(\xi) > k_{d}$ and $n_c(\tilde \xi) < q_c$ (Case 1), $n_c(\tilde \xi) = q_c$ (Case 2), and $n_d(\xi) = k_{d}$ (Case 3).

\textbf{Case 1:} Suppose $n_d(\xi) > k_{d}$ and $n_c(\tilde \xi) < q_c$. We show that $\xi-\chi_c^t \in \Xi^E$ and $\tilde{\xi}+\chi_c^t \in \Xi^E$.

$\xi-\chi_c^t$ and $\tilde{\xi}+\chi_c^t$ satisfying the capacity constraints and the exchange-feasibility constraints follows from the same arguments as in Case 1 in the proof of Proposition \ref{prop:convexbal}. Moreover, $\xi-\chi_c^t$ and $\tilde{\xi}+\chi_c^t$ satisfying the ceiling constraints and the floor constraints follows from the same arguments as in Case 1 in the proof of Proposition \ref{prop:convexdiv}.

\textbf{Case 2:} Suppose $n_c(\tilde \xi) = q_c$, then we argue that there exists $t' \neq t$ such that $\xi_c^{t'} < \tilde{\xi}_{c}^{t'}$. This follows from the facts that $n_c(\tilde \xi) = q_c \geq n_c(\xi)$ and $\xi_c^t>\tilde{\xi}_{c}^{t}$. We show that $\xi-\chi_c^t+\chi_{c}^{t'} \in \Xi^{DE}$ and $\tilde{\xi}+\chi_c^t-\chi_{c}^{t'}\in \Xi^{DE}$.

$\xi-\chi_c^t+\chi_{c}^{t'}$ and $\tilde{\xi}+\chi_c^t-\chi_{c}^{t'}$ satisfying the capacity constraints and the exchange-feasibility constraints follow
from the same arguments as in Case 2 in the proof of Proposition \ref{prop:convexbal}. Moreover, $\xi-\chi_c^t$ and $\tilde{\xi}+\chi_c^t$ satisfying the ceiling
constraints and the floor constraints follow from the same arguments as in Case 2 in the proof of Proposition \ref{prop:convexdiv}.

\textbf{Case 3:} Suppose $n_d(\xi) = k_{d}$, then, for this case, we first note that $n_d(\xi) \leq n_d(\tilde \xi)$. Next, we consider two subcases of this case: $n_c(\xi) \le n_c(\tilde \xi)$ (Subcase 3.1), and $n_c(\xi) > n_c(\tilde \xi)$ (Subcase 3.1).

\textbf{Subcase 3.1:} Suppose $n_c(\xi) \leq n_c(\tilde \xi)$. Then similar to Case 2 above, it follows that there exists $t' \neq t$ such that $\xi_c^{t'} < \tilde{\xi}_{c}^{t'}$. Then one can verify that $\xi-\chi_c^t+\chi_{c}^{t'} \in \Xi^{DE}$ and $\tilde{\xi}+\chi_c^t-\chi_{c}^{t'}\in \Xi^{DE}$.

\textbf{Subcase 3.2:} Suppose $n_c(\xi) > n_c(\tilde \xi)$, then we first observe that there exists a $c' \neq c$ with $d(c')=d$ such that
$n_{c'}(\xi) < n_{c'}(\tilde \xi)$. This is because, otherwise, we cannot have $n_d(\xi) \leq n_d(\tilde \xi)$. Next, we observe that $n_{c'}(\xi) < n_{c'}(\tilde \xi)$
implies that $ \xi_{c'}^{t'} <  \tilde{\xi}_{c'}^{t'}$ for some $t'$. We show that $\xi-\chi_c^t+\chi_{c'}^{t'} \in \Xi^{DE}$ and $\tilde{\xi}+\chi_c^t-\chi_{c'}^{t'}\in \Xi^{DE}$

First, we note that $\xi-\chi_c^t+\chi_{c'}^{t'}$ and $\tilde{\xi}+\chi_c^t-\chi_{c'}^{t'}$ satisfying the capacity constraints and the exchange-feasibility constraints follow from the same arguments as in Subcase 3.1 in the proof of Proposition \ref{prop:convexbal}. Finally, we argue the following:

\smallskip
\noindent \emph{Ceiling and floor constraints for $\xi-\chi_c^t+\chi_{c'}^{t'}:$} $\xi-\chi_c^t+\chi_{c'}^{t'}$ satisfies the ceiling and floor constraints since (i) $\xi_c^t>\tilde{\xi}_{c}^{t}$ implies $\xi_c^t>p_{c}^{t}$, hence $\xi_{c}^{t}(\xi-\chi_c^t+\chi_{c'}^{t'}) \geq p_{c}^{t}$, and (ii) $\xi_{c'}^{t'} < \tilde{\xi}_{c'}^{t'}$ implies $\xi_{c'}^{t'}<q_{c'}^{t'}$, hence $\xi_{c}^{t}(\xi-\chi_c^t+\chi_{c'}^{t'}) \leq q_{c}^{t}.$

\smallskip
\noindent \emph{Ceiling and floor constraints for $\tilde{\xi}+\chi_c^t-\chi_{c'}^{t'}:$} $\tilde{\xi}+\chi_c^t-\chi_{c'}^{t'}$ satisfies the ceiling and floor constraints since (i) $\xi_c^t>\tilde{\xi}_{c}^{t}$ implies $q_{c}^{t}>\tilde{\xi}_{c}^{t}$, hence $\xi_{c}^{t}(\tilde{\xi}+\chi_c^t-\chi_{c'}^{t'}) \leq q_{c}^{t}$, and (ii) $\xi_{c'}^{t'} < \tilde{\xi}_{c'}^{t'}$ implies $\xi_{c'}^{t'}>p_{c'}^{t'}$, hence $\xi_{c}^{t}(\tilde{\xi}+\chi_c^t-\chi_{c'}^{t'}) \geq p_{c}^{t}.$

Hence, we established that $\Xi^{DE}$ is M$^{\natural}$-convex.

The result then follows from Theorem \ref{thm:ttc} and Lemmas \ref{lem:cheb} and \ref{lem:policyobjective} because $\Xi^{DE}$ is M$^{\natural}$-convex.
\end{proof}
\medskip

\begin{proof}[Proof of Proposition \ref{prop:bal-ex}]

We show that the balanced-exchange policy \[\Xi^{B} = \left \{\xi \in \Xi^0 \mathlarger{\mathlarger{\mid}} \sum_{t,c:d(c)=d} \xi_c^t = k_{d} \text{ for all } d \in \cald \right \}\]
is M-convex; therefore, it is also M$^{\natural}$-convex.

For a distribution $\xi \in \mathbb N^{|\C|\times |\T|}$, let us call
\begin{enumerate}
\item $q_c \geq \sum_t \xi_c^t$ (equivalently, $q_c \geq n_c(\xi)$) the \emph{capacity constraint} for school $c\in \calc$, and
\item $\sum_{t,c:d(c)=d} \xi_c^t = k_{d}$  the \emph{balanced-exchange constraint} for district $d \in \cald.$
\end{enumerate}

Suppose that there exist $\xi,\tilde{\xi}\in \Xi^{B}$ and $(c,t) \in \calc \times \calt$ with $\xi_c^t>\tilde{\xi}_{c}^{t}$. Let us denote $d(c)$ by $d$.

Firstly, note that $n_d(\xi) = n_d(\tilde \xi) = k_{d}$. Next, we consider two cases: $n_c(\xi) \leq n_c(\tilde \xi)$ (Case 1) and $n_c(\xi) > n_c(\tilde \xi)$ (Case 2).

\textbf{Case 1:} Suppose $n_c(\xi) \leq n_c(\tilde \xi)$, then we argue that there exist $t' \neq t$ such that $\xi_c^{t'} < \tilde{\xi}_{c}^{t'}$. This follows from the facts that $n_c(\tilde \xi)  \geq n_c(\xi)$ and $\xi_c^t>\tilde{\xi}_{c}^{t}$. Then, both $\xi-\chi_c^t+\chi_{c}^{t'}$ and $\tilde{\xi}+\chi_c^t-\chi_{c}^{t'}$ clearly satisfy both the capacity constraints and the balanced-exchange constraints. This is because, $n_c$ and $n_d$ values among $\xi$ and $\xi-\chi_c^t+\chi_{c}^{t'}$ remain the same, and $n_c$ and $n_d$ values among $\tilde \xi$ and $\tilde{\xi}+\chi_c^t-\chi_{c}^{t'}$ remain the same. Thus, $\xi-\chi_c^t+\chi_{c}^{t'} \in \Xi^{B}$ and $\tilde{\xi}+\chi_c^t-\chi_{c}^{t'}\in \Xi^{B}$.

\textbf{Case 2:} Suppose $n_c(\xi) > n_c(\tilde \xi)$, then we first argue that there exists a $c' \neq c$ with $d(c')=d$ such that $n_{c'}(\xi) < n_{c'}(\tilde \xi)$.
This is because, otherwise, we cannot have $n_d(\xi) = n_d(\tilde \xi)$. Next, we observe that $n_{c'}(\xi) < n_{c'}(\tilde \xi)$ implies that $ \xi_{c'}^{t'} <  \tilde{\xi}_{c'}^{t'}$ for some $t'$. Then, we show that $\xi-\chi_c^t+\chi_{c'}^{t'} \in \Xi^{B}$ and $\tilde{\xi}+\chi_c^t-\chi_{c'}^{t'}\in \Xi^{B}$.

\smallskip
\noindent \emph{Balanced-exchange constraints:} Both $\xi-\chi_c^t+\chi_{c'}^{t'}$ and $\tilde{\xi}+\chi_c^t-\chi_{c'}^{t'}$ clearly satisfy the balanced-exchange constraints, as $n_d$ values among $\xi$ and $\xi-\chi_c^t+\chi_{c}^{t'}$ remain the same, and $n_d$ values among $\tilde \xi$ and $\tilde{\xi}+\chi_c^t-\chi_{c}^{t'}$ remain the same.

\smallskip
\noindent \emph{Capacity constraints for $\xi-\chi_c^t+\chi_{c'}^{t'}:$} $\xi-\chi_c^t+\chi_{c'}^{t'}$ clearly satisfies the capacity constraint for $c$ and it also satisfies the capacity constraint for $c'$ since $n_{c'}(\xi) < n_{c'}(\tilde \xi) \leq q_{c'}$. It also satisfies the capacity constraints for other schools since $\xi$ satisfies the capacity constraints for other schools.

\smallskip
\noindent \emph{Capacity constraints for $\tilde{\xi}+\chi_c^t-\chi_{c'}^{t'}:$} $\tilde{\xi}+\chi_c^t-\chi_{c'}^{t'}$ clearly satisfies the capacity constraint for $c'$ and it also satisfies the capacity constraint for $c$ since  $n_c(\tilde \xi) < n_c(\xi) \leq q_c$. It also satisfies the capacity constraints for other schools since $\tilde{\xi}$ satisfies the capacity constraints for other schools.

Thus, $\xi-\chi_c^t+\chi_{c'}^{t'} \in \Xi^{B}$ and $\tilde{\xi}+\chi_c^t-\chi_{c'}^{t'}\in \Xi^{B}$.

Hence, we established that $\Xi^{B}$ is M-convex and hence, M$^{\natural}$-convex.

The result then follows from Theorem \ref{thm:ttc} and Lemmas \ref{lem:cheb} and \ref{lem:policyobjective} because $\Xi^{B}$ is M$^{\natural}$-convex.
\end{proof}

\begin{proof}[Proof of Proposition \ref{prop:bal-mix}]
We show that the combination of school-level diversity and balanced-exchange policies  \[\Xi^{DB} = \left \{\xi\in \Xi^0 \mid q_c^t \geq \xi_c^t \geq p_c^t \text{ , } \forall (c,t)\in \calc \times \calt \text{ and }
  \sum_{t,c:d(c) = d} \xi_c^t = k_{d} \text{ , } \forall d \in \cald\right \}.\] is M-convex; therefore, it is also M$^{\natural}$-convex.

As in Propositions \ref{prop:convexdiv} and \ref{prop:bal-ex}, for a distribution $\xi \in \mathbb N^{|\C|\times |\T|}$, let us call
\begin{enumerate}
\item $q_c \geq \sum_t \xi_c^t$ (equivalently, $q_c \geq n_c(\xi)$) the \emph{capacity constraint} for school $c\in \calc$,
\item $q_c^t \geq \xi_c^t$ the \emph{ceiling constraint} for school $c\in \calc$ and type $t\in \calt$
\item $\xi_c^t \geq p_c^t$ the \emph{floor constraint} for school $c\in \calc$ and type $t\in \calt$, and
\item $\sum_{t,c:d(c)=d} \xi_c^t = k_{d}$  the \emph{balanced-exchange constraint} for district $d \in \cald.$

\end{enumerate}

Suppose that there exist $\xi,\tilde{\xi}\in \Xi^{DB}$ and $(c,t) \in \calc \times \calt$ with $\xi_c^t>\tilde{\xi}_{c}^{t}$. Let us denote $d(c)$ by $d$. Firstly, note that $n_d(\xi) = n_d(\tilde \xi) = k_{d}$. Next, we consider two cases: $n_c(\xi) \leq n_c(\tilde \xi)$ (Case 1) and $n_c(\xi) > n_c(\tilde \xi)$ (Case 2).

\textbf{Case 1:} Suppose $n_c(\xi) \leq n_c(\tilde \xi)$, then we argue that there exist $t' \neq t$ such that $\xi_c^{t'} < \tilde{\xi}_{c}^{t'}$. This follows from the facts that $n_c(\tilde \xi) \geq n_c(\xi)$ and $\xi_c^t>\tilde{\xi}_{c}^{t}$. Then, both $\xi-\chi_c^t+\chi_{c}^{t'}$ and $\tilde{\xi}+\chi_c^t-\chi_{c}^{t'}$ satisfying the capacity constraints and the balanced-exchange constraints follow from the same arguments as Case 1 in the proof of Proposition \ref{prop:bal-ex}. Finally, $\xi-\chi_c^t+\chi_{c}^{t'}$ and $\tilde{\xi}+\chi_c^t-\chi_{c}^{t'}$ satisfying the ceiling and floor constraints follow from the same arguments in Case 2 of Proposition \ref{prop:convexdiv}.

Thus, $\xi-\chi_c^t+\chi_{c}^{t'} \in \Xi^{DB}$ and $\tilde{\xi}+\chi_c^t-\chi_{c}^{t'}\in \Xi^{DB}$.

\textbf{Case 2:} Suppose $n_c(\xi) > n_c(\tilde \xi)$, we first argue that there exists a school $c' \neq c$ with $d(c')=d$ such that $n_{c'}(\xi) < n_{c'}(\tilde \xi)$. This is because, otherwise, we cannot have $n_d(\xi) \leq n_d(\tilde \xi)$. Next, we observe that $n_{c'}(\xi) < n_{c'}(\tilde \xi)$ implies that $ \xi_{c'}^{t'} <  \tilde{\xi}_{c'}^{t'}$ for some $t'$.

Then, both $\xi-\chi_c^t+\chi_{c'}^{t'}$ and $\tilde{\xi}+\chi_c^t-\chi_{c'}^{t'}$ satisfying the capacity constraints and the balanced-exchange constraints follow
from the same arguments as Case 2 in the proof of Proposition \ref{prop:bal-ex}. Finally, $\xi-\chi_c^t+\chi_{c'}^{t'}$ and $\tilde{\xi}+\chi_c^t-\chi_{c'}^{t'}$ satisfying the ceiling and floor constraints follow from the same arguments in Subcase 3.2 of Proposition \ref{prop:mix}.

Thus, $\xi-\chi_c^t+\chi_{c'}^{t'} \in \Xi^{DB}$ and $\tilde{\xi}+\chi_c^t-\chi_{c'}^{t'}\in \Xi^{DB}$.

Hence, we established that $\Xi^{DB}$ is M-convex. The result then follows from Theorem \ref{thm:ttc} and Lemmas \ref{lem:cheb} and \ref{lem:policyobjective}
because $\Xi^{DB}$ is M-convex, and, therefore, M$^{\natural}$-convex.
\end{proof}

\end{document}